\documentclass[%
 reprint,
showpacs,
 amsmath,amssymb,amsfonts,amsthm,amscd,bbm,
 aps,
prb,
]{revtex4-1}

\usepackage{graphicx}
\usepackage{dcolumn}
\usepackage{bm}

\newcommand{\muB}{\mu_{\rm B}}
\newcommand{\VD}{V_{\rm I}}
\newcommand{\VI}{V_{\rm C}}

\newif\ifNOSUP \NOSUPfalse

\begin{document}

\preprint{JO/RAR-QHHF}

\title{Manifestation of many-body interactions in the integer quantum Hall effect regime}

\author{Josef Oswald}
\email{Josef.Oswald@unileoben.ac.at}
\affiliation{%
Institut f\"{u}r Physik, Montanuniversit\"{a}t Leoben, Franz-Josef-Strasse 18, 8700 Leoben, Austria
}%

\author{Rudolf A.\ R\"{o}mer}
\email{R.Roemer@warwick.ac.uk}
\affiliation{
Department of Physics and Centre for Scientific Computing, University of Warwick, Coventry, CV4 7AL, UK
}%
\affiliation{
Department of Physics and Optoelectronics, Xiangtan University, Xiangtan 411105, Hunan, China
}

\date{$Revision: 1.2 $, compiled \today}

\begin{abstract}
We use the self-consistent Hartree-Fock approximation for numerically addressing the integer quantum Hall (IQH) regime in terms of many-body physics at higher Landau levels (LL). The results exhibit a strong tendency to avoid the simultaneous existence of partly filled spin-up and spin-down LLs. Partly filled LLs appear as a mixture of coexisting regions of full and empty LLs. We obtain edge stripes with approximately constant filling factor $\nu$ close to half-odd filling at the boundaries between the regions of full and empty LLs, which we explain in terms of the $g$-factor enhancement as a function of a locally varying $\nu$ across the compressible stripes.The many-particle interactions follow a behaviour as it would result from applying Hund's rule for the occupation of the spin split LLs. The screening of the disorder and edge potential appears significantly reduced as compared to screening based on a Thomas-Fermi approximation. For addressing carrier transport, we use a non-equilibrium network model (NNM) that handles the lateral distribution of the experimentally injected non-equilibrium chemical potentials $\mu$. 
 
\end{abstract}



\maketitle


\section{\label{sec:level1}Introduction}

The quantum Hall (QH) effect is a well-characterized example of complex quantum behavior emerging, in a two-dimensional (2D) solid-state system such as (mostly) doped semiconductors, due to the interplay of magnetic confinement, disorder and many-body interactions.\cite {KliDP80} A perpendicular magnetic field $B$ restricts the electronic charges to motion along circular orbits in the 2D plane, eventually leading to a Landau quantization when the cyclotron orbit $2 \pi l_c^2$, with magnetic length $l_c = \sqrt{\hbar/eB}$, enters a fully phase-coherent low-temperature regime, usually on microscopic length scales of $l_c \lesssim 100$ nm. Resistances and conductances are quantized as $R_{xy}= \frac{p}{q} \frac{h}{e^2}$ and $G= \frac{q}{p} \frac{e^2}{h}$ with $p=1$ and $q$ integer for the integer QH (IQH) and additionally $p$ an integer for the fractional QH (FHQ) effect.\cite{TsuSG82}  The composite fermion picture aims to explain the FQH effect via the many-body concept of quasi particles while in the IQH effect such interactions seem to play less of a central role.\cite{WeiK11} 
Longitudinal transport takes place along quasi one-dimensional (1D) directed channels, mainly along the edges of the sample, created at the intersections of the local Landau levels (LLs) and the Fermi energy.\cite{Hal82} Disorder stabilizes the plateaus in $R_{xy}$ via electronic localization, while $R_{xx}\neq 0$ only between plateaus. Non-interacting models, such as the Chalker-Coddington network model,\cite {ChaC88} can explain much of this IQH phenomenology and are excellent in characterizing the universal properties of plateau-to-plateau transitions.

However, interactions cannot be completely ignored even for IQH physics. The exchange interaction is known to lead to an enhanced $g$-factor for spin splitting.\cite{Jan69,RN440,NicHKW88,UshNHF90,Nomura2013a,Vionnet2016a} Recent scanning gate experiments, investigating edge stripes passing quantum point contacts at high $B$,\cite{PasRIE14} indicate  modified screening behavior within the compressible stripes. Furthermore, a very recent investigation of the local nature of compressibility in the bulk of a IQH sample strongly challenges existing single-particle theories.\cite{KenSKO17}
%
%

A major step towards modelling electron-electron interactions in the QH is due to Chklovskii, Shklovskii and Glazman (CSG) who considered screening via the electrostatics of the edge channel region.\cite{ChkSG92} The key feature of their model is the different screening capabilities of partly and fully filled LLs that appear in alternating order towards the edge depletion zone of the electronic system. CSG predict wide stripe-like regions of partly filled LLs that can get up to hundreds of nanometers wide. These compressible stripes are separated by usually much narrower in-compressible stripes consisting of completely filled LLs. The essence of the CSG picture is that the CSs screen out almost the entire slope of the bare electrostatic potential near the sample boundaries and thus generate terraces in the uprising edge potential while the electron density $n$ and the local filling factor $\nu$ change continuously across the CS.
Charge propagation is governed by the CSs, although the CSG approach does not detail the mechanism of the 1D quantized transport.

Additional many-body effects such as, e.g.\ exchange interactions, are not included in the CSG approach. Nevertheless the observed exchange-induced spin-splitting of LLs\cite{FogS95,RN440,Nomura2013a,Vionnet2016a} can be modelled phenomenologically by introducing an effective $g$-factor enhancement.\cite{Nomura2013a} Theoretical descriptions\cite{RN440,RN439} of the enhancement predict a characteristic $\nu$ factor dependence of almost vanishing enhanced $g$ at even integer $\nu$ and maximal enhancement at odd integer $\nu$. 
Here we show, based on a (converged) self-consistent Hartree-Fock approach coupled with a non-equilibrium network to model transport in a QH bar set-up, how locally resolved terraces of completely filled LLs are being populated when $n$ and $B$ change across QH plateaus. While transport is in good qualitative and quantitative agreement with IQH physics, our locally resolved $\nu(\mathbf{r})$ data show clear distinctions to the CSG picture. We find intriguing features in the locally resolved $\nu$ with enhanced half-odd $\nu$ values and coinciding with regions of charge transport. A behaviour reminiscent of Hund's rule leads to the avoidance of partially filled spin-up and spin-down LLs in the same spatial region. Reinterpreting our Hartree-Fock results using the language of enhanced $g$-factors allows us to recover the aforementioned phenomenology.\cite{OswR17} Here, we compare the Hartree-Fock results to Hartree and non-interacting calculations and also present higher temperature effects. 

\section{\label{methods}Methods}

\subsection{Self-consistent Hartree-Fock in Landau basis}
\label{sec-hartree-fock}

In order to model a high-mobility heterostructure in the QH regime, we
consider a 2DES in the $(x,y)$-plane subject to a perpendicular magnetic
field $\vec{B} = B\vec{e}_z$ described by the
Hamiltonian 
\begin{equation}
H^\sigma_{\rm 2DES} =
           h^\sigma + \VI =
           \frac{(\vec{p}-e\vec{A})^2}{2m^*} +
           \frac{\sigma g^* \muB B}{2} +
           \VD(\vec{r}) +
           \VI(\vec{r},\vec{r}'),
\label{eq-hamiltonian}
\end{equation}
where $\sigma = \pm 1$ is a spin degree of freedom, $\VD$ is a smooth
random potential modeling the effect of the electron-impurity
interaction, $\VI$ represents the electron-electron interaction term and
$m^*$, $g^*$, and $\mu_{\rm B}$ are the effective electron mass,
$g$-factor, and Bohr magneton, respectively.
In order to avoid edge effects we impose a torus geometry of size
$L\times L$ onto the system.\cite{YosHL83} The electron-impurity
interaction is modeled by an electrostatic potential due to a remote
impurity density separated from the plane of the 2DES by a spacer-layer
of thickness $d$, as found for instance in modulation-doped GaAs-GaAlAs
heterojunctions. Within the plane of the 2DES, this creates a random,
spatially correlated potential with a typical length scale $d$. We use
$N_{\rm I}$ Gaussian-type "impurities", randomly distributed at
$\vec{r}_s$, with random strengths $w_s \in [-W,W]$, and a fixed width
$d$. \cite{SohR07}
The areal density of impurities is given by $n_{\rm I} = N_{\rm I}/L^2$. 
The Coulomb interaction potential will be parametrized by $\gamma$ to allow us to continually adjust the
interaction strength; $\gamma=1$ corresponds to the bare Coulomb
interaction.
For the system's many-body state, $|\Phi\rangle$, we use the usual
ansatz\cite{Aok79,MacA86} of an anti-symmetrized product of single
particle wave-functions, which we choose as a linear combination of Landau states\cite{SohR07}
$\psi_\alpha^\sigma(\vec{r}) = \sum_{n=0}^{N_{\rm LL}-1}
  \sum_{k=0}^{N_\phi-1} \vec{C}^{\alpha,\sigma}_{n,k} \chi_{n,k}(\vec{r})
$, 
with $N_{\rm LL}$ being the number of LLs and the periodic
Landau functions $\chi_{n,k}(\vec{r})$.
The number of flux quanta
piercing the 2DES is given by $N_\phi=L^2/2\pi
l_{\rm c}^2$, yielding a total number of $M = N_{\rm LL} N_\phi$ states
per spin direction. The filling of the system is characterized by the
filling factor $\nu = N_{\rm e}/N_\phi$, with $N_{\rm e}$ the
number of electrons in the system and areal density $n_{\rm e} = N_{\rm
  e}/L^2$. In terms of $N_{\mathrm{e},\uparrow}$ spin-up and $N_{\mathrm{e},\uparrow}$ spin-down electrons, we can hence write $\nu= \nu_\uparrow + \nu_\downarrow$ and $N_\mathrm{e}= N_{\mathrm{e},\uparrow} + N_{\mathrm{e},\downarrow}$. The total LL density is given by $n_0 = eB/h$ and $l_{\rm c} = \sqrt{\hbar/eB}$ the magnetic length.
%
%
A variational minimization of $\langle\Psi|H_{\rm 2DES}|\Psi\rangle$
with respect to the coefficients $\vec{C}^{\alpha,\sigma}_{n,k}$ \cite{Aok79,YosF79,MacA86,MacG88} yields the self-consistent Hartree-Fock-Roothaan
equation, \cite{Roo51}
$
\mathbf{H}^\sigma\mathbf{C}^\sigma = \mathbf{C}^\sigma\mathbf{E}^\sigma
\label{eq-hfe}
$, 
with
$\mathbf{C}^\sigma = (\vec{C}^{\sigma}_1,\dots,\vec{C}^{\sigma}_M)$
the matrix of eigenvectors and
$\mathbf{E}^\sigma =
\mbox{diag}(\epsilon^\sigma_1,...,\epsilon^\sigma_M)$
the diagonal matrix of the eigenvalues $\epsilon^\sigma_1 \le
\epsilon^\sigma_2 \le \dots \le \epsilon^\sigma_M$.
Following the Aufbau principle, \cite{CanL00} the density matrix is
constructed starting from the energetically lowest lying state up to the
Fermi level $\epsilon_{\rm F}$. In our calculations, we keep $N_{\rm e}$
fixed and compute $\epsilon_{\rm F}$ as the energy of the highest
occupied state afterwards.
%
We start the self-consistency process using solution of the non-interacting Hamiltonian $\mathbf{h}^\sigma=\langle nk |h^\sigma|n'k'\rangle$ as initial guess for the coefficients $\mathbf{C}^{\sigma}$. From this solution,
$\mathbf{C}^{(0)}$, we construct the density and Fock matrices and
finally the full Hamiltonian. \cite{SohR07} Diagonalization yields an improved
solution, $\mathbf{C}^{(1)}$. The process continues until convergence of
the density matrix has been achieved. \cite{SohR07} In all results present here,
convergence of the HF scheme is computed by a modified Broyden mixing\cite{Sri84,Joh88,Uiberacker2012a} and achieved when the accuracy $\varepsilon \le 10^{-6}$.

\subsection{The non-equilibrium network model (NNM)}
\label{CS-sec-oswaldmodel}

The NNM describes the spatial distribution of the experimentally injected \emph{non-equilibrium} chemical potentials, $\mu(x,y)$, of 2D electron systems in the high magnetic field regime\index{High magnetic field regime}.\cite{Osw16} Differences to the equilibrium CCN \cite{ChaC88} and similarities with the models proposed in Refs. \onlinecite{PolS95,DamHM11} have been discussed previously.\cite{OswR15,Osw16}
In the NNM, the local backscattering function $P$ is given as
\begin{equation}
P(E_\mathrm{LL}; E_\mathrm{F}, \ell, U) = \exp \left[ -\frac{\ell^2 (E_{\rm F}-E_{\rm LL})}{e U} n_0 \right],
\label{CS-eq-12}
\end{equation}
where $E_{\rm LL}$ is the saddle energy that corresponds to the LL center, $E_{\rm F}$ represents the Fermi energy, and $\ell$ and $U$ are connected to the Taylor expansion of the involved SP\index{Saddle
point}: $\ell$ is the period and $U$ the amplitude of a 2D
cosine-potential, which has the same 2nd order Taylor expansion like
the actual saddle potentials. This version of a QH network has been demonstrated to be quite successful for a variety of transport simulations of realistic sample structures and  experimental setups.\cite{RN26, RN68, Uiberacker2012a}
The backscattering function can be rewritten using the filling factor formulation   
\begin{equation}
P (\nu; \Gamma, \ell, U) = \exp \left [ - \Gamma \frac{ \ell^2 \sqrt{\pi} (\nu-\lfloor \nu \rfloor -\frac{1}{2}) }{eU} n_0 \right ] ,
\label{fill}
\end{equation}
which allows seamless integration to the self-consistent Hartree-Fock approach of section \ref{sec-hartree-fock} (here $\lfloor \nu \rfloor$ denotes the integer value of $\nu$). We note that Eq.\ \eqref{fill} can be derived from Eq.\ \eqref{CS-eq-12} by assuming a Gaussian shaped DOS of width $\Gamma$. Due to the randomness of the potential fluctuations $\Gamma$ will be significantly larger than $eU$, which represents a typical single saddle. If estimating $\Gamma$ as $\approx 3 eU$ and $\ell \approx 100$ nm as the order of magnitude for the mean fluctuation period, for moderate $B\approx 3$ T, the argument of the exponential will become $\approx - (\nu - \lfloor \nu \rfloor - 0.5)/0.025$.\cite{OswO06}
When taking into account Thomas-Fermi screening on the basis of a simple Hartree interaction, Eq.\ \eqref{fill} is equivalent to Eq.\ \eqref{CS-eq-12}. However, if aiming at many body interactions the Eqs.\ \eqref{CS-eq-12} and \eqref{fill} are no longer equivalent, because exchange effects are not captured by the Hartree potential in Eq.\ \eqref{CS-eq-12}, while they are well included in the local carrier density profile and local $\nu$ that is used in Eq.\ \eqref{fill}. 

For the application of the NNM to transport in a many particle quantum system it is important that no local quantities such as a local conductivity or local Ohm’s law are used by the NNM, because that would imply the (forbidden) possibility to establish the path of the carriers while moving from one current contact to the next. Currents are calculated at the designated current contacts only as a post-processing step, that is, after obtaining the self consistent solution. Potential differences are taken from the voltage probes, which can be defined in principle at arbitrarily chosen locations of the network. Then resistances, such as Hall resistance and longitudinal resistance, are computed depending on the arrangement of the chosen contact pairs as in real experiments.

\section{\label{results}Results}

\begin{figure}[tb]
\includegraphics[width=\columnwidth]{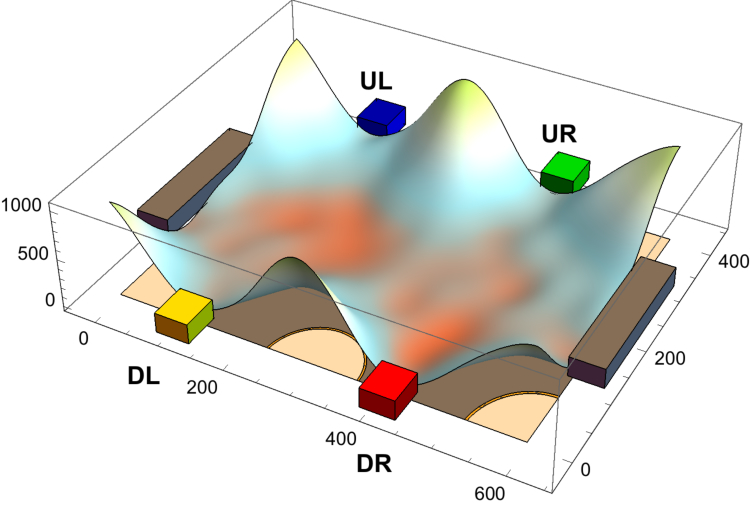}
\caption{\label{bare_pot} Bare potential for a Hall bar of size $600 \times 400$ nm$^2$. An \emph{edge confinement} potential is created by repulsive Gaussian peaks of $1$ V amplitude at the corners and in the middle of the longitudinal sides, leaving openings for current contacts at the ends (long grey bars) and voltage probes (blue, green, yellow and red cuboids labelled UL, UR, DL, DR, resp.) along the sample. An additional potential \emph{disorder} of maximally $\pm 10$ mV is generated by randomly distributed Gaussians (with $N_\text{I}=200$, $d=40$ nm and $w_s=4$ mV). This disorder is coded onto the surface of the total potential as small height fluctuations. The bottom plane indicates the depletion regions in light color while the electron-rich area is dark.}
\end{figure}

\begin{figure*}[tb]
(a)\includegraphics[width=0.44\textwidth]{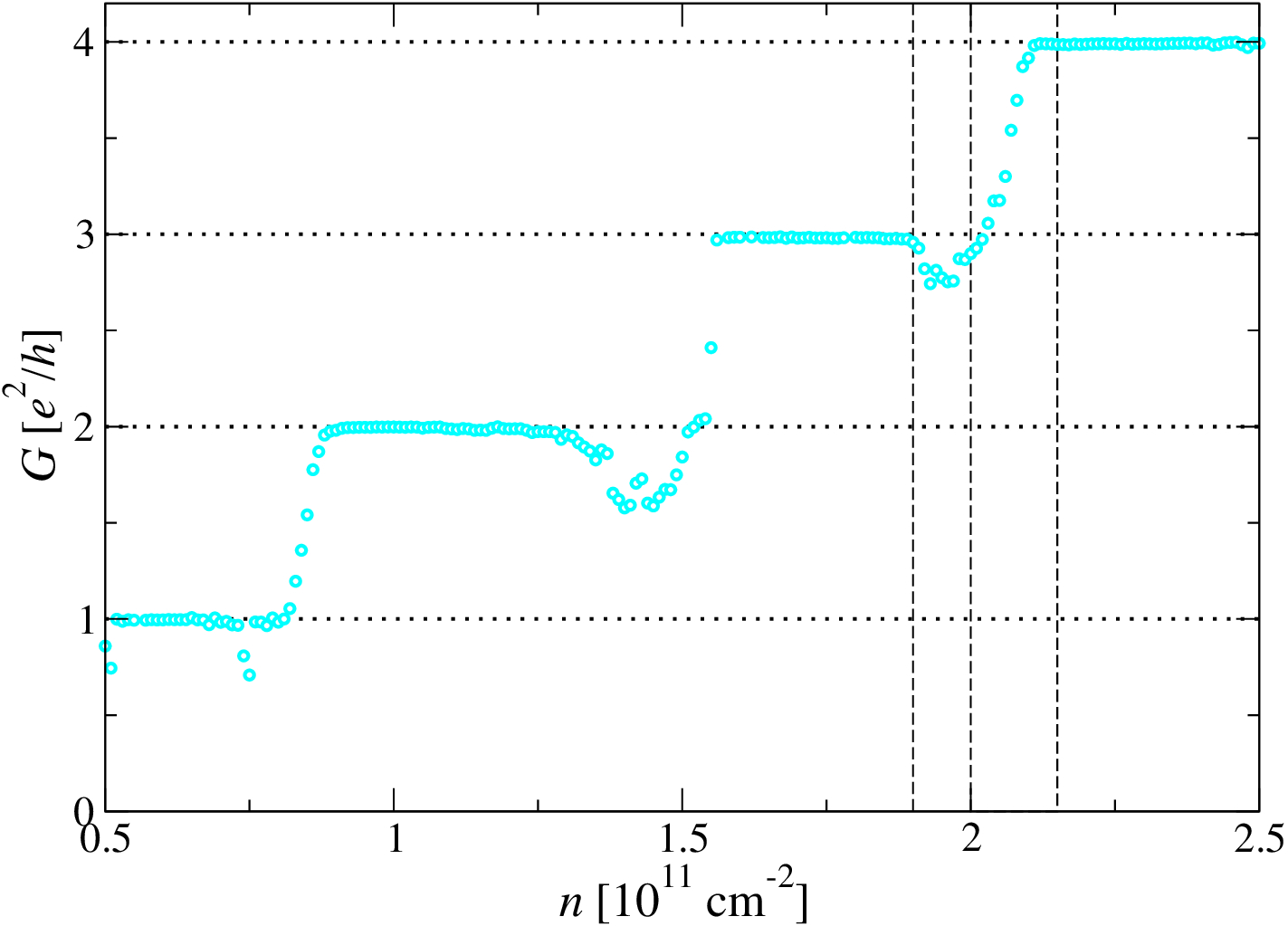}
(b)\includegraphics[width=0.45\textwidth]{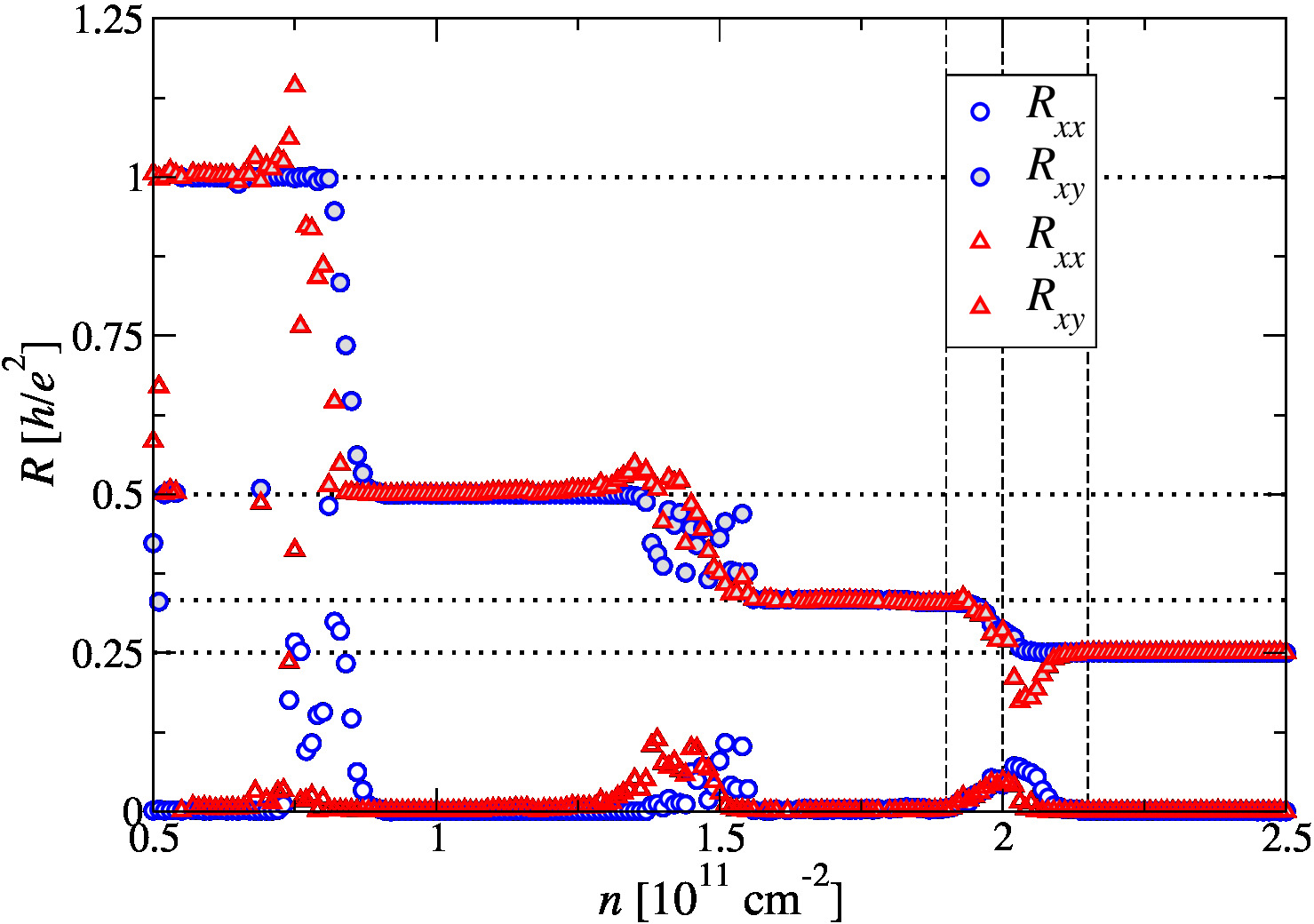}
\caption{
\label{fig-conductance-resistance-density}
(a) Two-point conductance $G$ versus carrier density $n$ at fixed magnetic field of $B=3$ T. The horizontal dotted lines indicate integer multiples of $e^2/h$.
(b) $R_{xx}$ and $R_{xy}$ versus $n$ at fixed $B=3$ T. The open and filled symbols represent the data of different UL/UR/DL/DR contact pairs according to Fig.~\ref{bare_pot}. Horizontal dotted lines indicate $h/e^2 m$ for $m=1, 2, 3, 4$. The vertical dashed lines indidate the three density values of $n= 1.9, 2.0, 2.15$ ($\times 10^{11}$ cm$^{-2}$).}
\end{figure*}

\begin{figure*}[tb]
(a)\includegraphics[width=0.6\textwidth]{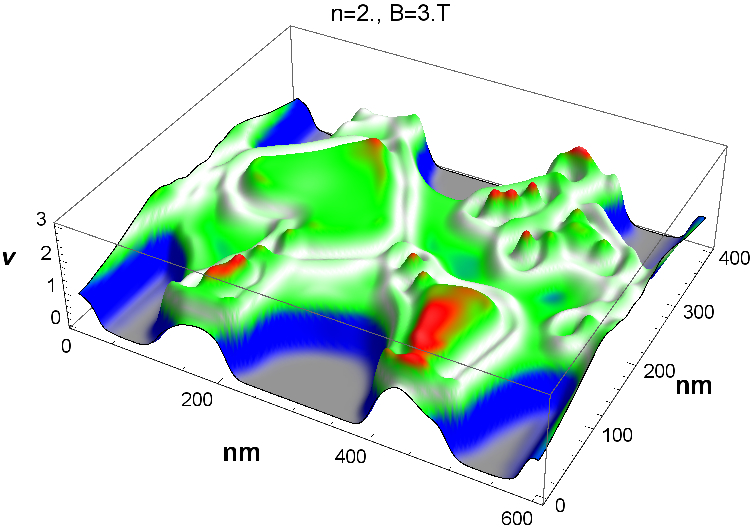} 
\begin{minipage}{0.32\textwidth}\vspace*{-50ex}
(b)\includegraphics[width=\textwidth]{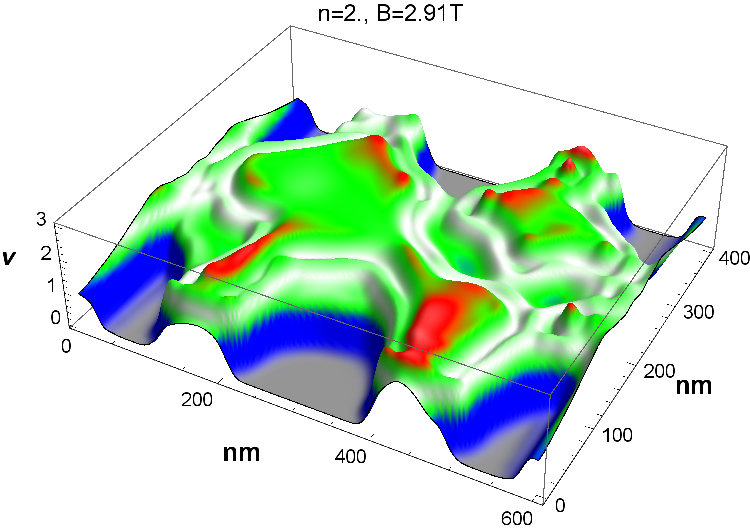}\\ 
(c)\includegraphics[width=\textwidth]{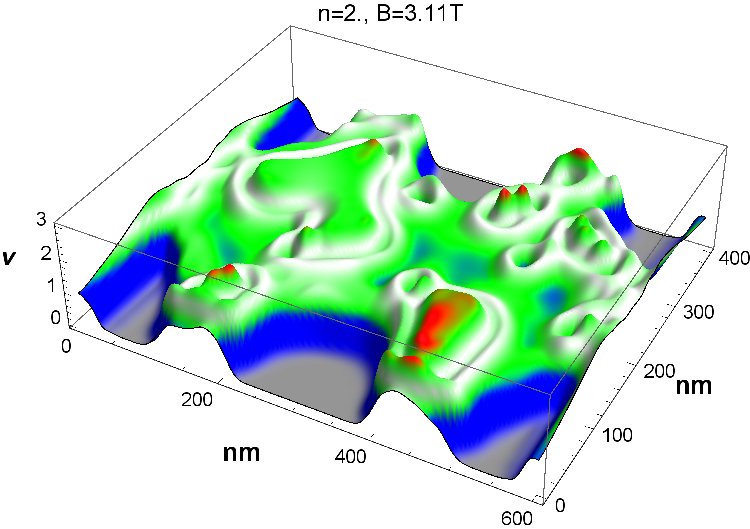}
\end{minipage}
(d)\includegraphics[width=0.3\textwidth]{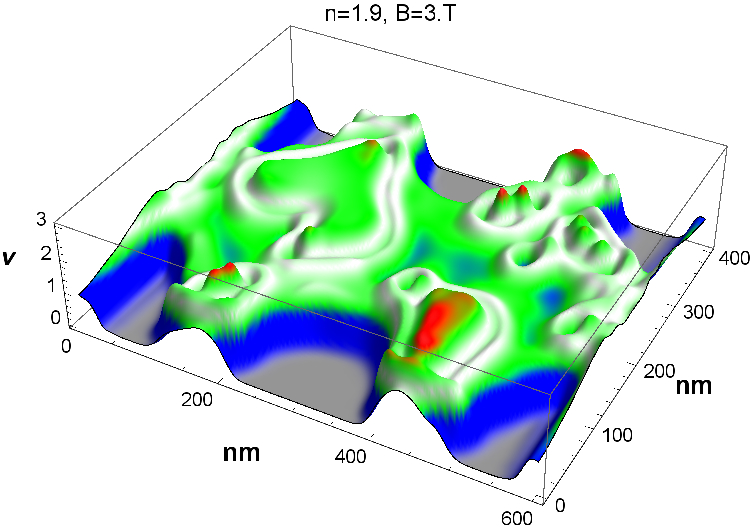}
(e)\includegraphics[width=0.3\textwidth]{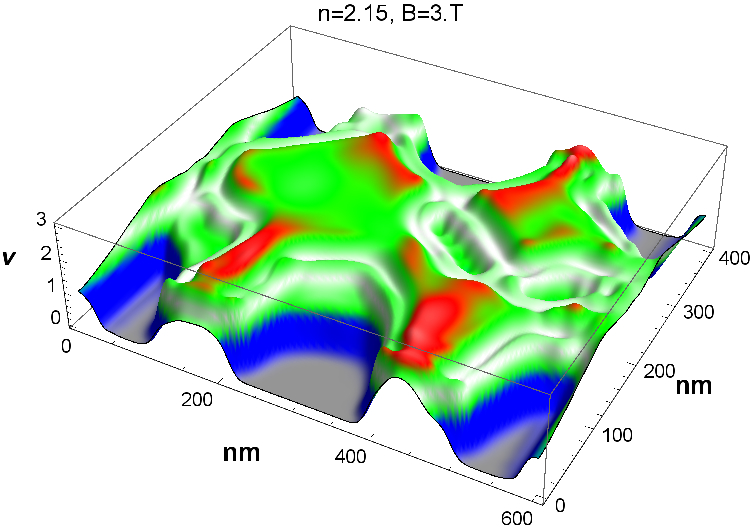}
(f)\includegraphics[width=0.3\textwidth]{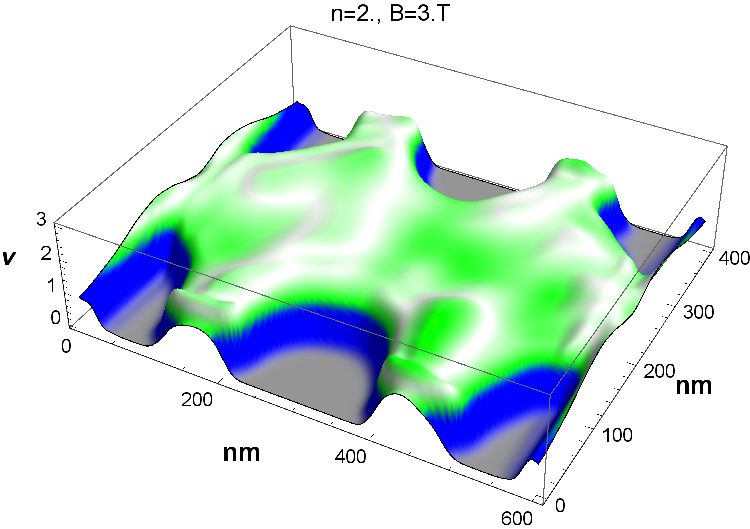}
\caption{
\label{fig-fillingfactor-up} Spatial filling factor distribution of $\nu_\uparrow$ for the highest partly filled LL close to the $\nu = 3 \rightarrow 4$ plateau transition at $T=1K$ 
(a) with $B=3$ T and $n=2\times 10^{11}$ cm$^{-2}$ of the carrier density sweep in Fig.~\ref{fig-conductance-resistance-density} and also the magnetic field sweep in Fig.\ 2 of Ref.\ \onlinecite{OswR17}; 
(b) corresponds to the beginning of the  $\nu = 4 \rightarrow 3$ transition at $B=2.91$ T while 
(c) is at the end of the transition at $B=3.11$ T for the magnetic field sweep.\cite{OswR17}
Panel (d) corresponds to the carrier density sweep in Fig.~\ref{fig-conductance-resistance-density} with $n=1.9 \times 10^{11}$ cm$^{-2}$ while 
(e) is at $n=2.15\times 10^{11}$ cm$^{-2}$. 
Last, (f) has $B=3$ T and $n=2\times 10^{11}$ cm$^{-2}$ as in (a) but at temperature $T\approx 20$K ($2$m$e$V).
The colors represent the filling factor, where blue means the first LL for $\nu_\uparrow = 0 \rightarrow 1$, green the second LL for $\nu_\uparrow = 1 \rightarrow 2$ and red the third. The filling factor range close to $\nu_\uparrow = 1.5$  is highlighted in light gray in order to identify the stripes appearing close to the half filled top LL. Corresponding results for $\nu_\downarrow$ are shown in Fig.\ \ref{fig-fillingfactor-down}.}
\end{figure*}

\begin{figure*}[tb]
(a)\includegraphics[width=0.6\textwidth]{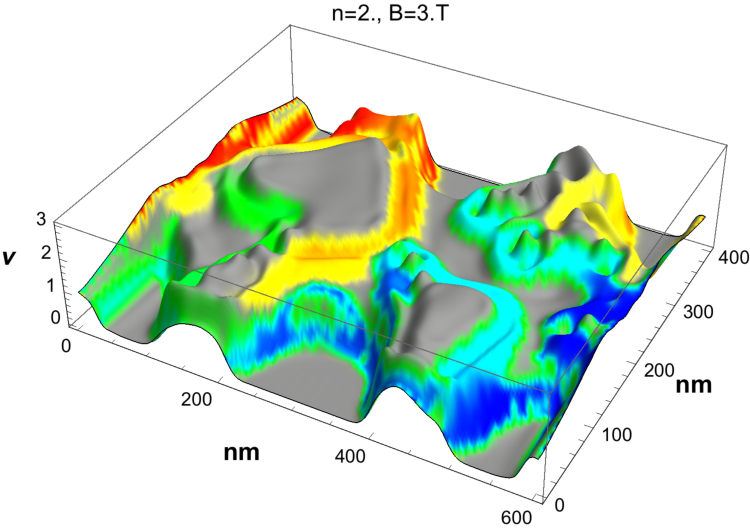}
\begin{minipage}{0.32\textwidth}\vspace*{-50ex}
(b)\includegraphics[width=\textwidth]{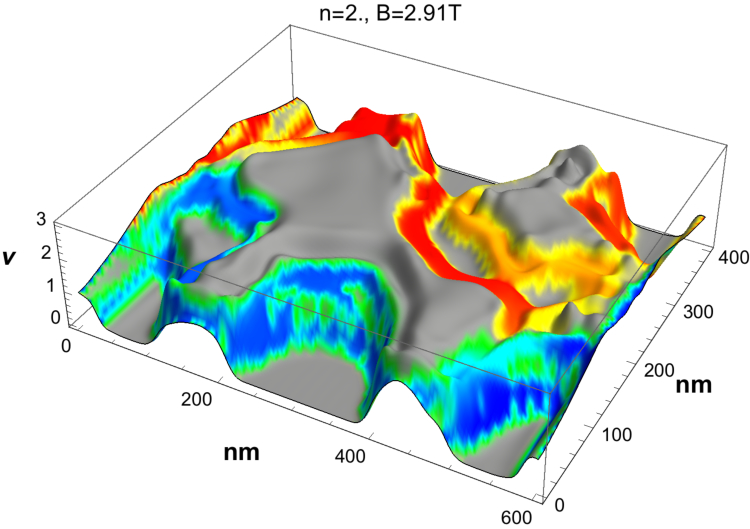}\\
(c)\includegraphics[width=\textwidth]{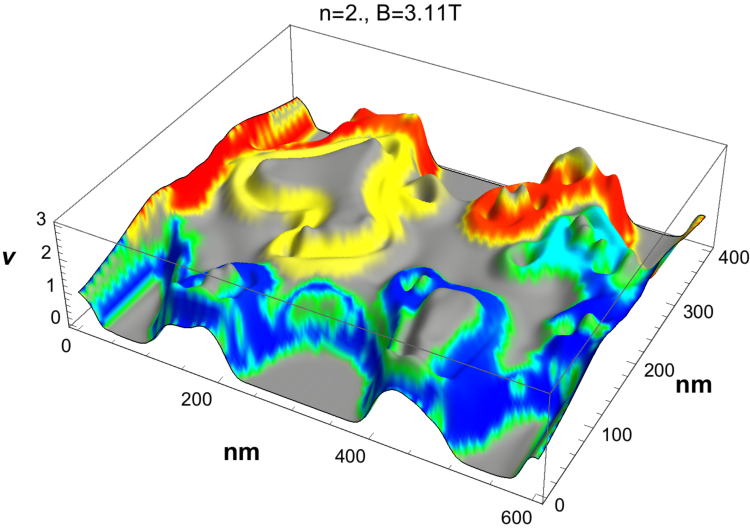}
\end{minipage}
(d)\includegraphics[width=0.3\textwidth]{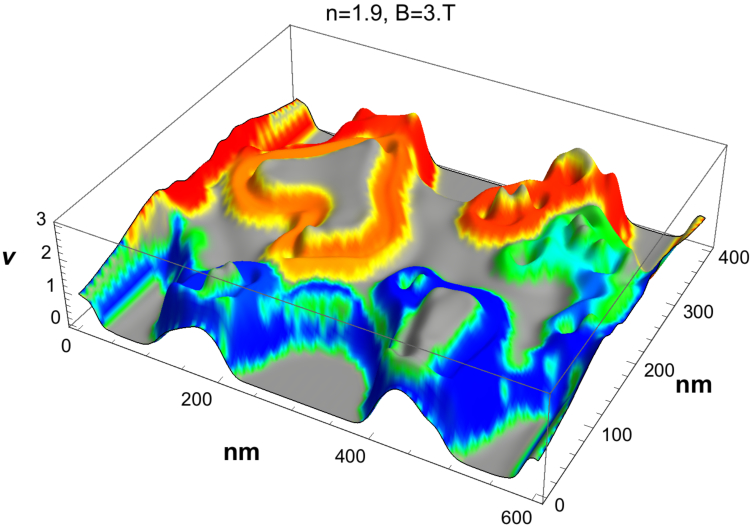}
(e)\includegraphics[width=0.3\textwidth]{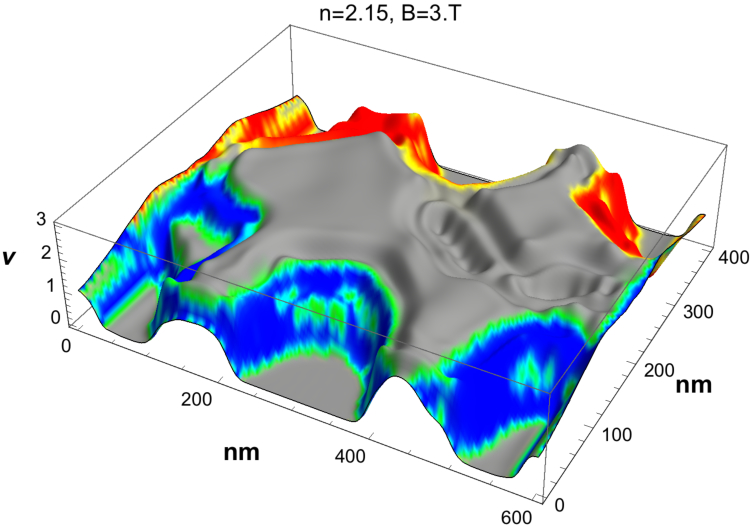}
(f)\includegraphics[width=0.3\textwidth]{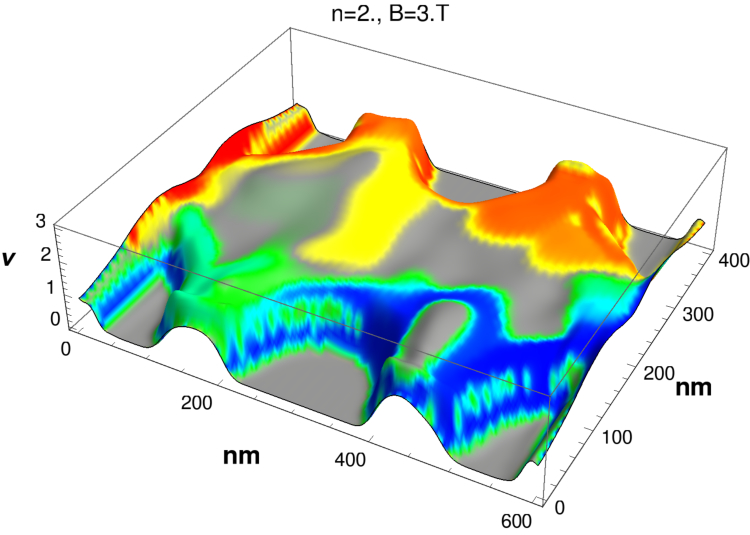}
\caption{
\label{fig-chemicalpotential-up} Spatial distribution of non-equilibrium chemical potential $\mu$ (colors) shown on top of the corresponding $\nu_\uparrow$ distribution (grey heights) as in Fig.\ \ref{fig-fillingfactor-up}. The colors represent $\mu$ in arbitrary units with overall clock-wise propagating potential reducing from the high potential supplied to the current contact on the left from red to orange and yellow while the low potential is supplied to the current contact on the right and is indicated as increasing from blue to cyan to green. Panels (a), (b), \ldots, (f) have the same parameters as in Fig.\ \ref{fig-fillingfactor-up}, such that (b), (a) and (c) show a change of $B$ field from $2.91$ T to $3$ T and $3.11$ T, resp., at constant carrier density $n=2\times 10^{11}$ cm$^{-2}$ while in (d), (a) and (e) $n$ changes from $1.9\times 10^{11}$ cm$^{-2}$ to $2\times 10^{11}$ cm$^{-2}$ and on to $2.15\times 10^{11}$ cm$^{-2}$, resp., at constant $B=3$ T. Panel (f) corresponds to $B=3$ T and $n=2\times 10^{11}$ cm$^{-2}$ as in (a) but at temperature $T\approx 20$K ($2$m$e$V). Corresponding results for $\nu_\downarrow$ are shown in Fig.\ \ref{fig-chemicalpotential-down}.
}
\end{figure*}

Figure \ref{bare_pot} shows the model potential of a Hall bar structure of total size of $600 \times 400$ nm$^2$ that gets filled with $480$ electrons, corresponding to a nominal average carrier density of $n = 2 \times 10^{11}$ cm$^{-2}$. We note that this density corresponds to an effective density of $\approx 2.5 \cdot 10^{11}cm^{-2}$ in the "bulk" region of the “structured” Hall sample of Fig.\ \ref{bare_pot}. Correspondingly, the transport data presented in the following appears shifted from $n$ according to the effective density.

The self-consistent solution of the Hartree-Fock calculation of $\nu(x,y)$ for different $B$ and $n$ is sampled at $186 \times 126$ spatial positions, i.e.\ with $3.226$ nm resolution in longitudinal ($x$) and $3.175$ nm in transversal ($y$) direction. For the range of $B$ values considered, this allows good resolution of structures below $l_c$, which ranges from $l_c\approx 14$ nm at $B=3.3$ T to $l_c \approx 17$ nm at $B=2.25$ T. The $\nu(x, y)$ values are then transferred to the NNM which calculates the self-consistent lateral distribution of the experimentally injected non-equilibrium electrochemical potentials $\mu(x, y)$.
We emphasize that the resulting $\mu$ is assumed \emph{not} to act back on $\nu$, which otherwise would introduce an additional nested self-consistence loop. Physically, this means that our transport calculations represent the linear response at vanishing small excitation close to thermal equilibrium, as is the case also in most of the transport experiments in the quantum Hall regime far from the QH breakdown regime. 

\subsection{\label{transport}Transport}

In Ref.\ \onlinecite{OswR17}, we have shown that the $B$-dependences of the longitudinal resistance, $R_{xx}$, and the Hall resistance, $R_{xy}$, exhibit the expected $R_{xx}$ peaks in the transition regime between QH plateaus and Hall plateaus at zero $R_{xx}$. In the transition regime between plateaus (cp.\ Fig.\ 2 of Ref.\ \onlinecite{OswR17}) we observe strong resistance fluctuations in both, as also expected for QH structures of mesoscopic size.\cite{Simmons1991}
In Fig.\ \ref{fig-conductance-resistance-density}, we show additionally the $n$-dependence of the transport data, obtained quite similar to the $B$-field dependence described above. The two-point conductance, $G_{xx}$, is shown in  Fig.\ \ref{fig-conductance-resistance-density} (a) while Fig.\ \ref{fig-conductance-resistance-density} (b) indicates $R_{xx}$ and $R_{xy}$ for two different contact pairs. 
The plateau transitions are again accompanied by strong fluctuations, which are clearly visible in the resistance data. The over- and undershoots in the $G_{xx}$ data can be understood to result from inhomogeneities due to the long-range disorder at mesoscopic size of the electron system. These become most influential in the vicinity of the plateau transitions.  
We emphasize that, overall, the quantization of the transport data is very well resolved and follows the expected $e^2/h$ multiples (and inverses thereof). For comparison, in Fig.\ \ref{fig-Gxx_ns_00-HH-HF}, we show how this picture is modified when instead of the full Hartree-Fock calculation, we use either just the Hartree or, indeed, the single-particle approach while keeping all other parameters unchanged. It is interesting to note that the full Hartree-Fock calculation seems to lead to more stable plateaus. 

\subsection{\label{filling}Spatial distribution of $\nu$}

Fig.\ \ref{fig-fillingfactor-up} shows the lateral $\nu_\uparrow$ distribution of the top spin-up LL during the $ \nu = 4 \rightarrow 3$ plateau transition. The spin-up level is higher in energy than the spin-down level and therefore depletion from $ \nu = 4 \rightarrow 3$ happens within the second spin-up LL only, while for spin-down we have filling factor $ \nu_{\downarrow} =2$ during the whole $ \nu = 4 \rightarrow 3$ plateau transition (cp. supplemental Fig.\ \ref{fig-fillingfactor-down}).
Fig.\ \ref{fig-fillingfactor-up} (a) is taken at $B=3$ T and show the situation in right at the transition. The figure sequence (b) $\rightarrow$ (a) $\rightarrow$ (c) in Fig.\ \ref{fig-fillingfactor-up} corresponds to a change in $B$ from $2.91$ T to $3.11$ T.
We see that at $B=2.91$ T the initially completely filled second spin-up LL starts to break up into sub-regions of filling factor $ \nu_{\uparrow} =2$, while there appear also sub-regions of filling factor $ \nu_{\uparrow} =1$. This gives an average total filling factor between $ \nu =4$ and $ \nu =3$ (Fig.\ \ref{fig-fillingfactor-up} (b)). At $B=3$ T the clusters for $ \nu_{\uparrow} =2$ have shrunk further (see Fig.\ \ref{fig-fillingfactor-up} (c)) and finally at $B=3.11$ T they just represent isolated droplets, while the region of filling factor $ \nu_{\uparrow} =1$ already dominates.

The behaviour is quite similar for the change of density shown in the sequence (d) $\rightarrow$ (a) $\rightarrow$ (e) in Fig.\ \ref{fig-fillingfactor-up}, but of course in the opposite direction. At low density (Fig.\ \ref{fig-fillingfactor-up} (d)) the region for $\nu_{\uparrow} =1$ dominates and the clusters of $ \nu_{\uparrow} =2$ are just isolated droplets. In the middle of the plateau transition, where the regimes of the $B$ field sweep and the density sweep cross each other at  $\nu_{\uparrow} \approx 1.5$ for the spin-up electrons (spin-down electrons remain at $ \nu_{\downarrow} =2$, which gives in total $\nu =3.5$), we find that about half of the area is covered by the $\nu_{\uparrow} =1$ and half by the $ \nu_{\uparrow} =2$ clusters. At larger carrier density the $\nu = 3 \rightarrow 4$ plateau transition is almost completed, exhibiting a domination of the area covered by the $ \nu_{\uparrow} =2$ region that starts to build a completely filled $\nu_{\uparrow} =2$ spin-up LL (Fig.\ \ref{fig-fillingfactor-up} (e)). This looks quite similar to the $\nu$ distribution at the beginning of the magnetic field sweep in Fig.\ \ref{fig-fillingfactor-up} (b). In addition, at the boundaries between clusters of $\nu_{\uparrow} =2$ and $\nu_{\uparrow} =1$ there appear terraces of almost constant filling factor close to $ \nu_{\uparrow} =1.5$ (see Figures \ref{fig-fillingfactor-up} (a)--(e)).

\subsection{\label{chemicalpotential}Spatial distribution of $\mu$}
In Fig.\ \ref{fig-chemicalpotential-up} we show the lateral distribution of $\mu$ for $B$ and $n$ values identical to the ones used in Fig.\ \ref{fig-fillingfactor-up}. The color coded values of $\mu$ are drawn onto the surfaces of the lateral $\nu$ distribution. It is clearly visible that in the bulk region, the non-equilibrium potential $\mu$ lies mainly along the half-odd integer features, i.e. $\nu_\uparrow=3/2$. The transmitting channels of the lower LLs appear directly at or close to the edge at the boundaries of the electron system. If the potentials mix at some locations this generates dissipation as is the case in the transition regime of the IQHE. While this happens only weakly in Figs.\ \ref{fig-chemicalpotential-up} (b), (c) and Figs.\ \ref{fig-chemicalpotential-up} (d), (e), this mixing appears to be quite strong in Fig.\ \ref{fig-chemicalpotential-up} (a), which represents a case close to the maximum of the $R_{xx}$-peak in Fig.\ \ref{fig-conductance-resistance-density} (b). We also observe dissipation at the voltage contacts. This is most clearly pronounced in Fig.\ \ref{fig-chemicalpotential-up}  (a). We note that the channels appear to merge into the metallic contact region of the voltage probes and reappear with changed magnitude. 

\subsection{\label{temperature}Local $\nu$ and $\mu$ at $T\approx 20$ K}

The $\nu$ and $\mu$ distributions as well as the transport data shown thus far have been computed at temperature $T=1$ K. In Figs.\ \ref{fig-fillingfactor-up} (f) and \ref{fig-chemicalpotential-up} (f), we have repeated the calculations, but now for $T= 2$ m$e$V ($\approx 20$ K). As can be seen, for $\nu_\uparrow$, the previously well-defined regions of constant filling factors dissolve into a much smoother density profile. The features at half-odd integer $\nu$ seem to be missing entirely. Correspondingly, the $\mu$ distribution appears smoother as well and has lost some structure details. 

\subsection{\label{HH-HF-00}Hartree and single-particle calculation}

\begin{figure*}[tb]
(a)\includegraphics[width=0.3\textwidth]{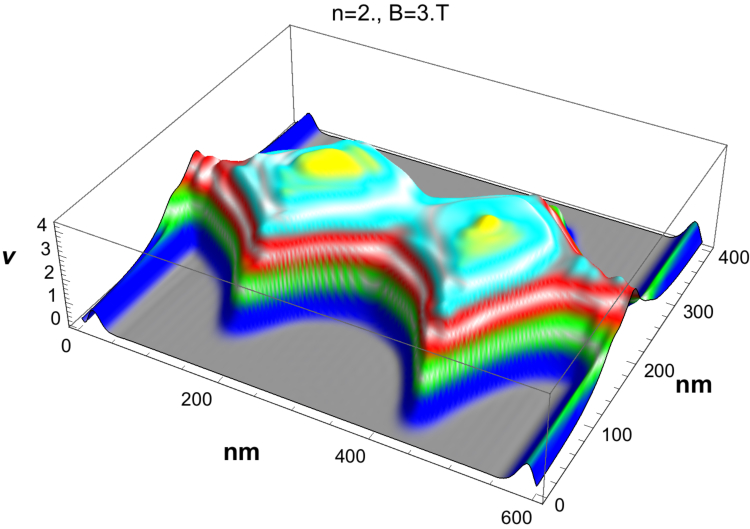} 
(b)\includegraphics[width=0.3\textwidth]{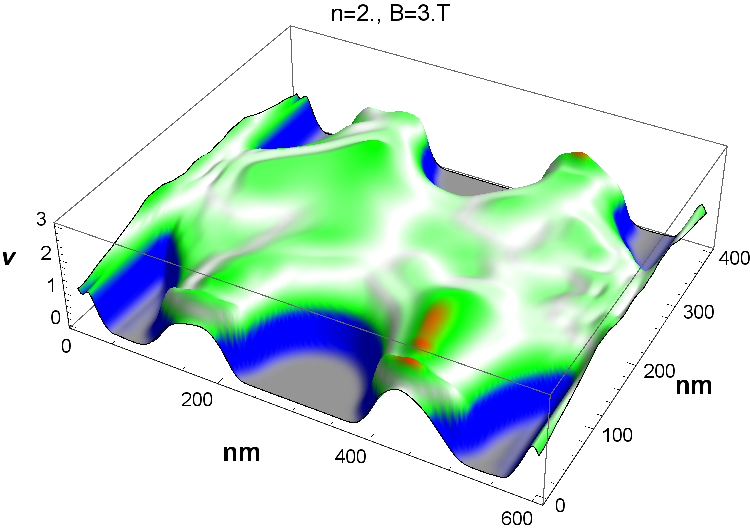} 
(c)\includegraphics[width=0.3\textwidth]{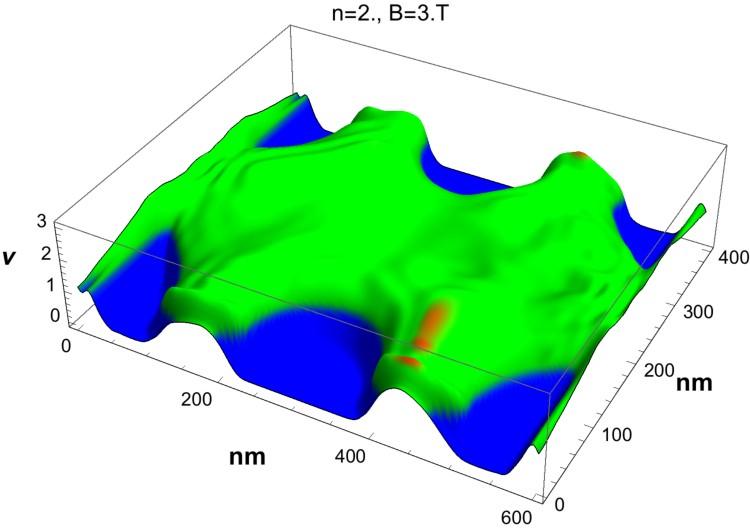} 
\caption{\label{CDS1-n200-00-HH-HF}\label{fig-00-HH-HF-up} Local $\nu_{\uparrow}$ distribution (a) based on the interaction-free single particle approximation (b) the Hartree approximation for $n=2 \times 10^{11}$ cm$^{-2}$ and $B=3.0$ T, which  corresponds to the $\nu = 3 \rightarrow 4$ plateau transition of Fig.~\ref{fig-conductance-resistance-density}. Panel (c) shows the same Hartree approximation as (b) but without color-highlighting of the half-odd integer filling.
The colors represent $\nu_{\uparrow}$ as in Fig.\ \ref{fig-fillingfactor-up} with light blue and yellow denoting LLs $4$ and $5$. The filling factor range close to $\nu_\uparrow = $ half-odd integer is highlighted in light gray from LL2 onwards in order to identify possible stripes appearing close to the half filled top LL.
Panel (c) shows the the same Hartree approximation as (b) but without color-highlighting the half-off integer filling.
}
\end{figure*}

Thus far we have implied that the results shown in Figs.\ \ref{fig-fillingfactor-up} and \ref{fig-chemicalpotential-up} (as well as the supplemental Figs.\ \ref{fig-fillingfactor-down} and \ref{fig-chemicalpotential-down}) are characteristic of the exchange-physics inherent in the Hartree-Fock interaction. In order to validate that hypothesis, we show in Fig.\ \ref{CDS1-n200-00-HH-HF} for $B=3$ and $n=2\times 10^{-11}$cm$^2$ the $\nu_\uparrow$ distribution for (a) a non-interacting and (b) a purely Hartree-interacting systems (cp.\ also Fig.\ \ref{ECCD-n200-00-HH-HF_spX} for $\nu_\downarrow$ and $\mu$ in the supplement).
For the interaction-free case, we find that the carriers are much confined in the centre of the bulk region. With Hartee interaction, the carriers  spread out laterally much more while the local $\nu_\uparrow$ at fixed $N_e$ in the centre reduces as compared to the interaction-free model. This is easily understood from effective screening of the edge potential in the Hartree calculation. We note that neither the non-interacting nor the Hartree calculation produces any of the half-odd $\nu_\uparrow$ features in Fig.\ \ref{fig-fillingfactor-up}. The Hartree results for $\nu_\uparrow$ retain some of the overall features of the Hartree-Fock behaviour for $\nu_\uparrow$, but at half-odd $\nu_\uparrow$, they show only a smooth rise in agreement with the Thomas-Fermi screening arguments of CSG. This lack of half-odd integer features is particularly visible when replotting Fig.\ \ref{CDS1-n200-00-HH-HF} (b) without any half-odd highlighting as in Fig.\ \ref{CDS1-n200-00-HH-HF} (c). 
We recall that a similar comparison for the transport data was already presented in Ref.\ \onlinecite{Osw16}.

\section{\label{discussion} Discussion}

\subsection{\label{CSG}Differences to CSG}

From the $n$ and $B$ sequences in Fig.\ \ref{fig-fillingfactor-up} one can see how the population/de-population of the spin-up LL works. Instead of getting an overall increase or decrease of the carrier density, we get shrinking or growing clusters of fully filled spin-up LL at $ \nu_{\uparrow} =2$ and growing or shrinking areas of depleted spin-up LL at $ \nu_{\uparrow} =1$. On average this results in a continuous change of the spatially averaged $\nu_\uparrow$. Therefore a combined $\nu_{\uparrow} =1.5$ is made up by half of the area taken up by clusters of $\nu_{\uparrow}=1$ and the other half taken up by $\nu_{\uparrow} =2$. The same happens subsequently if the spin-down LL gets de-populated at the transition $ \nu = 3 \rightarrow 2$ (not shown), which stays at $ \nu_{\downarrow} =2$ while populating/de-populating the top spin-up LL. 
This behaviour is different to the CSG model and suggests also that even a half-filled LL may provide only poor screening when compared to CSG's Thomas-Fermi-like continuous $n$ variation across the sample area. 
In addition, in our Hartree-Fock approach, we observe in Fig.\ \ref{fig-chemicalpotential-up} terraces of almost constant filling factor close to $\nu_{\uparrow} =1.5$ at the boundaries between clusters of $ \nu_{\uparrow} =2$ and $ \nu_{\uparrow} =1$ which create transmitting channels in the NNM. The width of the half-odd integer stripes appears to be of the order of $l_c$. We furthermore find \emph{two} parallel stripes of width $l_c$ at the boundaries between clusters $\nu_{\uparrow} =3$ and $\nu_{\uparrow} =2$ (cp.\ supplemental Fig.\ \ref{fig-fillingfactor-down-highLL}). This suggest that the origin of the stripes lies in the spatial dependence of Landau states similar, perhaps, to what is observed for the local density of states.\cite{HasCFS12}
The cluster boundaries and the boundaries of the fully filled LLs at the sample edge are the only regions where the many-particle electron system can exchange carriers close to equilibrium as in low excitation magneto-transport experiments. As a consequence, these boundaries are experimentally observed as transport channels.\cite{Osw16} While CSG suggest narrow, so called incompressible, stripes of fully filled LLs between the compressible stripes, in our case almost the whole space is divided into clusters of empty or full spin-up or spin-down LLs. If these clusters finally merge they add or remove an edge stripe to the sample boundary, which completes the plateau transition from $G =  3 e^2/h \leftrightarrow 4 e^2/h$.

\subsection{\label{half-odd}Features at half-odd $\nu$ and an effective Hund's rule}

An intriguing result is the appearance of half-odd integer terraces in the density profiles of the Hartree-Fock calculation. This is clearly different to the CSG model because they replace the terraces in the electrostatic edge potential. The other observation is that the electrons condense to droplets of either fully filled or empty spin-up or spin-down LLs which in higher LLs are separated by the half-odd integer terraces. This suggests that the electron system tries to avoid as much as possible the existence of partly filled spin-up and spin-down LLs in the same spatial region. This is similar to a Hund’s rule behaviour. Simply speaking, the higher spin-up states get pushed up in energy in order to keep on adding, as long as possible, further electrons with parallel spin to the lower spin-down state. In this way the occupation of the upper spin-up level gets delayed as compared to an occupation that follows just the energy of the pure Zeeman splitting, even if the cluster size already extends into regions of classically forbidden elevated potentials. This is effectively an enhancement of the spin-splitting energy and thus an effective $g$-factor enhancement.  The exchange-enhanced $g$-factor appears as a built-in effect without any need to introduce it by hand or even needing to think about its existence at all. However, in order to compare with single particle models and discuss the results in terms of single electrons like done in context with the CSG approach, we also have to introduce a $g$-factor, which appears to be considerably enhanced as compared to the bare electronic $g$-factor in order to meet the obtained results.

\subsection{\label{$g$-factor}Exchange-enhanced $g$-factor}

\begin{figure*}[tb]
(a)\includegraphics[width=0.95\columnwidth]{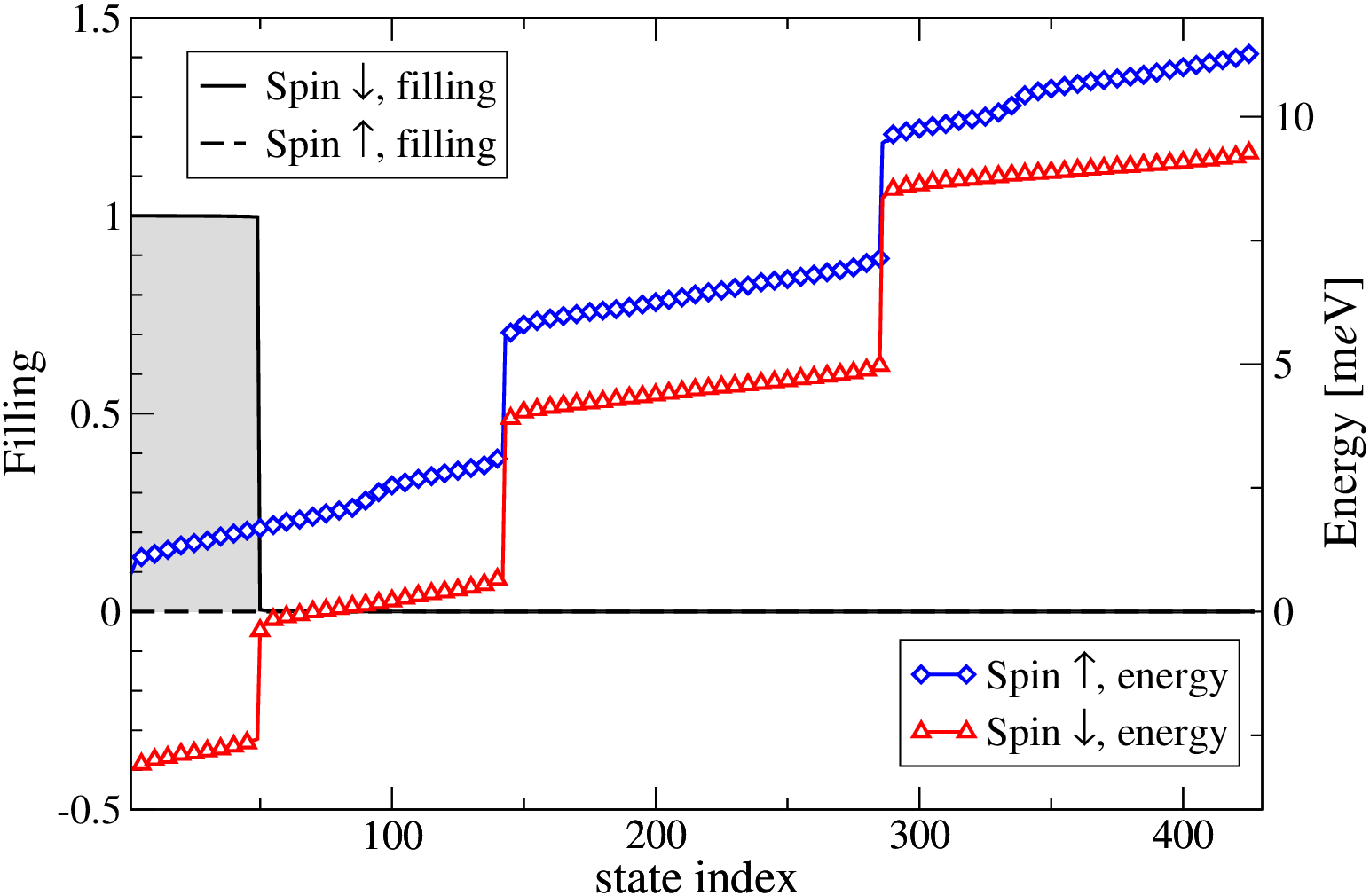}
(b)\includegraphics[width=0.95\columnwidth]{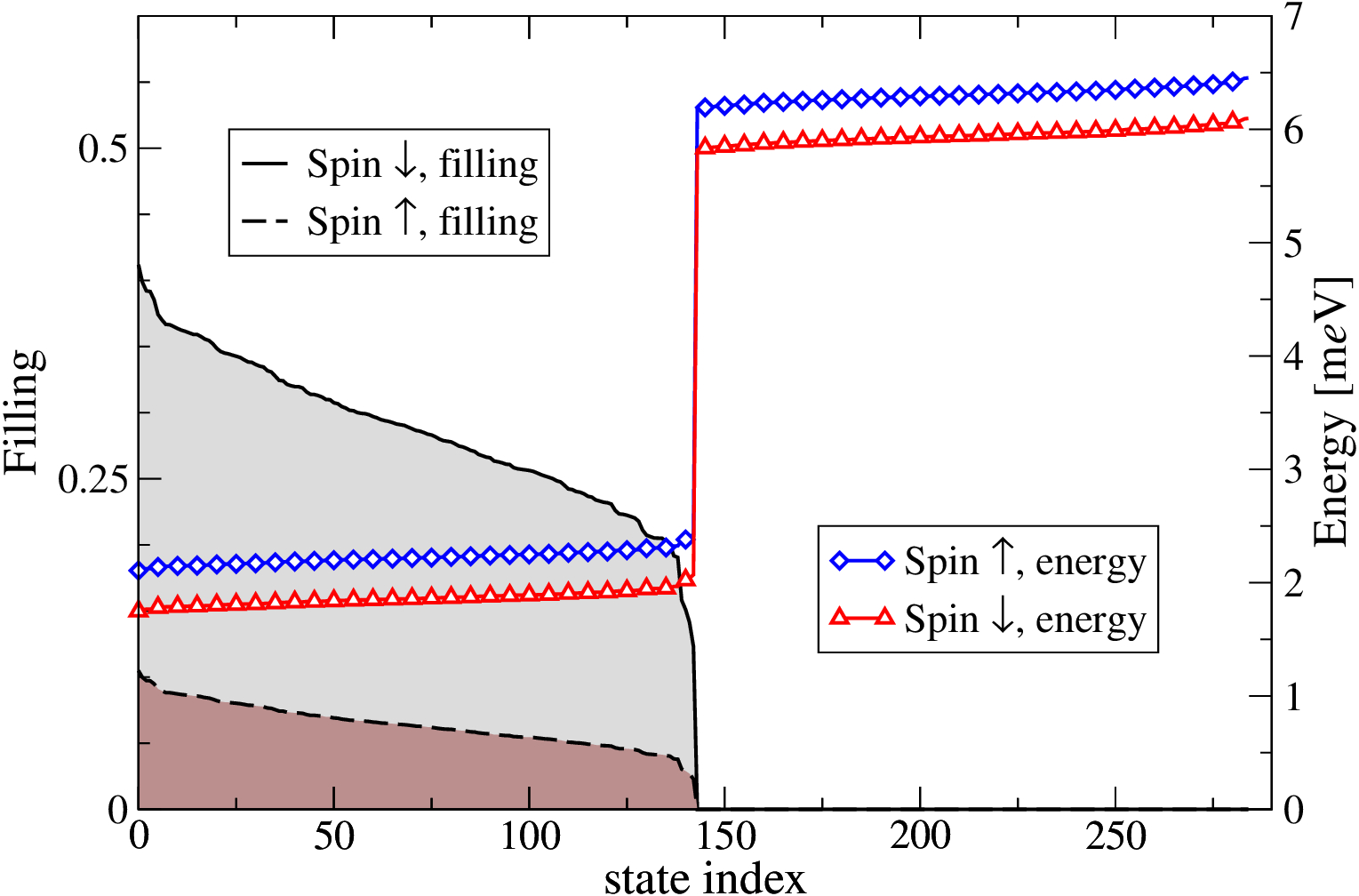}
\caption{
\label{fig-spectrum}
Energy spectrum ($\diamond$, $\triangle$) and filling (shaded lines)  for (a) Hartree-Fock and (b) Hartree interaction at filling factor $\nu=0.35$ and $B=2.36$ T. Spin up and down associations are as indicated in the legends. Only every $5$th symbol is shown for the energy spectrum.}
\end{figure*}

\begin{figure}\includegraphics[angle=0,keepaspectratio=true,width=0.45\textwidth]{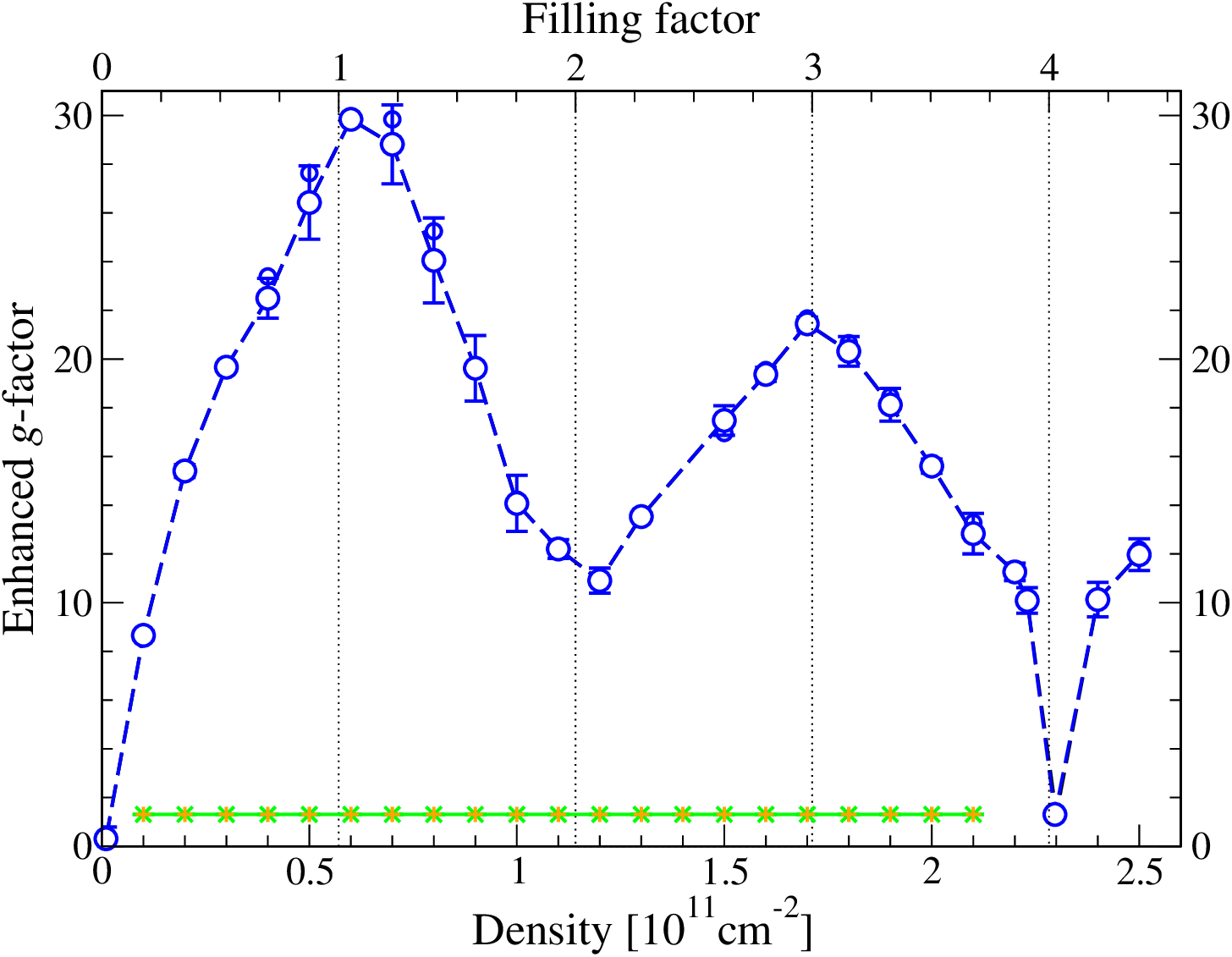}
\caption{\label{fig-enhanced-g}  
Plot of the $g$-enhancement factor (alternatively $\Delta E$) as a function of $n$ (and $\nu$). The open (blue $\circ$) circles corresponds to the Hartree-Fock calculation and denote the mean and error bars indicate the standard error averaging about at most $100$ energy differences close to the Fermi energy. Crosses (green $\times$) show corresponding results for the Hartree calculation. The (blue) dashed and (green) solid lines are a guide to the eye only. In both cases, $B=2.36$ T as in Fig. \ref{fig-spectrum}.
}
\end{figure}

Let us now discuss our findings using the language of an enhanced $g$-factor. In this way, we can make contact with the single-particle picture by including the majority of the many-body physics in the renormalized $g$. We have studied the energy splitting $\Delta E = g\ g_\mathrm{spin} \mu_B B$ where $\mu_{\rm B}$ denotes the Bohr magneton,   $g_\mathrm{spin}\approx 2$ is the bare electron $g$-factor, and $g$ its enhancement factor, respectively. This numerical study has been done for a $500 \times 500$ nm$^2$ test structure with a weak disorder potential of about $\pm 1.5$ m$e$V at $B=2.36$ T at different filling factors ranging from $\nu = 0$ to $4.3$. Averaging over $\Delta E$ between the occupied spin-down and the unoccupied spin-up states of Fig.\ \ref{fig-spectrum} (a), we can compute $g$ as shown in Fig.\ \ref{fig-enhanced-g}.  We can clearly see, that the occupied spin-down states and the empty spin-up states get pushed apart in energy. 
We find that there is indeed a considerable enhancement with $g \gg 1$ at odd $\nu$ while the enhancement drops for even $\nu$. The oscillatory behavior of the enhanced $g$-factor is already well known from the literature\cite{UshNHF90,NicHKW88} and had been recently investigated experimentally as well as theoretically. \cite{Nomura2013a} In the paper of Nomura et al.\ they find the same oscillatory behaviour. However, their investigation was done at $B< 1$ T and hence they observe a lower $g$-enhancement. In our simulations it seems that the mechanism driving the creation of the half-odd integer stripes gets weaker at $B \ll 2$ T.  
Fig.\ \ref{fig-spectrum} (b) details the situation for a purely Hartree-interacting system and, as shown in Fig.\ \ref{fig-enhanced-g}, the "enhancement" for this situation is $g \approx 1$. 

\subsection{\label{screening}Dynamic screening in the IQH regime}

\begin{figure*}[tb]
\includegraphics[width=\textwidth]{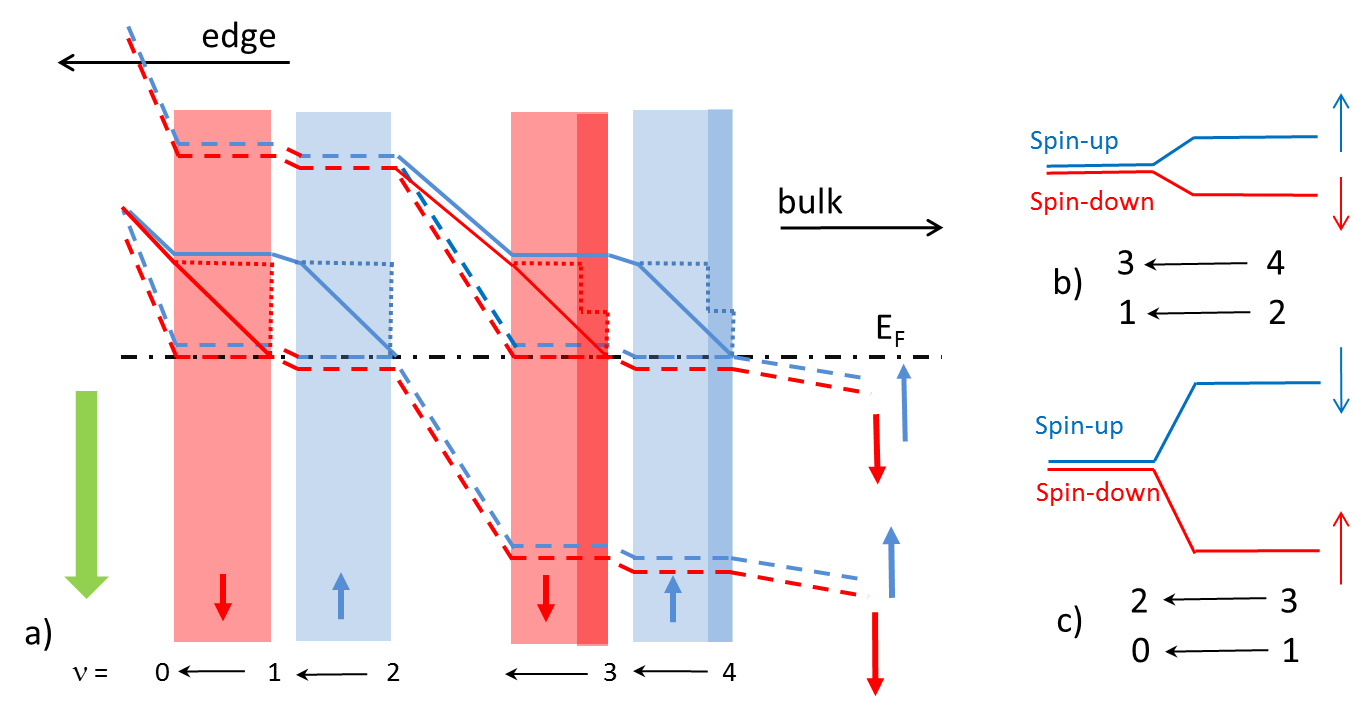}
\caption{\label{pushup}  (a) Schematic sketch of exchange initiated modifications for the CSG model: The screened edge potential according to CSG for spin-up is shown in dashed blue and for spin-down in dashed red; the bold lines indicate the exchange driven level shift for spin-up in blue and spin-down in red; The blue shaded bar indicates the CS of the partly filled spin-up level and the red shaded bar indicates the CS of the partly filled spin-down level, while the light shaded area indicates the original width due to CSG, while the narrow dark shaded area indicates the width after carrier redistribution due to exchange effects. The dotted lines schematically indicate the final spin-splitting after self consistent carrier redistribution.  (b) schematic spin splitting at even filling factor, the arrows indicate the tendency of exchange driven increase of Zeeman energy while the filling factor changes from even towards odd $\nu = 4 \rightarrow 3$ or $\nu = 2 \rightarrow 1$ that manifests also in the opening up of the spin-splitting indicated by the bold lines in a) while crossing a blue shaded CS; (c) schematic spin splitting at odd filling factor, the arrows indicate the tendency of decreasing the exchange driven  Zeeman energy if the filling factor changes from odd towards even $\nu = 3 \rightarrow 2$ or $\nu = 1 \rightarrow 0$ that manifests also in the closing of the spin-splitting indicated by the bold lines in a) while crossing a red shaded CS.}
\end{figure*}

In Fig.\ \ref{pushup} we attempt to explain our results by considering a locally varying exchange-enhanced $g$-factor.  In Fig.\ \ref{pushup} (a) the dashed lines schematically represent the edge potentials due to CSG. These resemble terraces at the position of the CS as indicated by the lightly shaded vertical bars. The inner stripe is of spin-up and the outer stripe of spin-down type in the chosen $B$ field direction. The red and blue solid lines indicate, how level energies should be modified if adding the effect of $g$-factor enhancement due to the changing local $\nu$. Starting at the inner (right) boundary  of the inner spin-up stripe, the local $\nu$ value is an even integer, resulting in		$g$-enhancement as indicated in Fig.\ \ref{pushup} (b). When crossing the CS towards its left edge, we approach an odd integer local $\nu$, which leads to maximal $g$-enhancement. Without carrier re-arrangement within the CS, the spin-up LL would get pushed up relative to the spin-down LL as indicated by the blue bold line in Fig.\ \ref{pushup} (a). Of course this cannot happen without loosing immediately all carriers in the stripe and subsequent carrier rearrangement. Consequently we have an effective edge potential as the sum of the electrostatic edge potential according to CSG and the varying exchange-enhanced spin splitting. Taken together, this determines the carrier distribution: The locally varying spin splitting strongly counteracts the pure electrostatic CSG screening and therefore the screening of the electrostatic part of the potential is suppressed to some extend. The uprising effective edge potential (blue bold line) creates an almost step-like change of the carrier distribution to the next lower odd-integer filling factor, leaving only a narrow feature of the order of the magnetic length. The jump in $\nu$ causes a self-consistently induced jump of the effective potential as indicated schematically by the blue dotted line. This in turn strongly reduces the degree of freedom for the carrier distribution, restricting the effective screening of the electrostatic potential. 

When we next cross the spin-down stripe, the cycle starts over again, but at odd $\nu$ with large spin splitting on the right of the red-shaded CS. Large spin splitting implies a low spin-down level relative to the spin-up level as indicated in Fig.\ \ref{pushup} (c). The levels get pushed towards each other while approaching the next even $\nu$ at the left side of the spin-down stripe. In this way the blue and red bold lines come close together again at the left boundary while crossing the spin-down stripe as shown in Fig.\ \ref{pushup} (a). 
The up-rising spin-down level (red solid line) abruptly looses all carriers and $\nu$ jumps to the next lower, even integer filling. Again a step-like change in the effective potential is initiated and the narrow half-odd integer feature remains at this step as before. This interpretation is consistent with the observed half-odd integer terraces in $\nu$ at the boundaries for spin-up and spin-down clusters. 

We note that an experimental indication of such a suppression of screening was recently reported by Pascher et al.\cite{PasRIE14} They investigated the screening of edge stripes passing a quantum point contact by scanning gate microscopy and found that their experimental results are not well described by 
Thomas-Fermi screening. 


\section{\label{outlooksummary}Summary and Outlook}

We employ a self-consistent Hartree-Fock approximation in higher LLs, studying screening and the lateral carrier distribution. 
Transport has been modeled by using a non-equilibrium network model.\cite{OswO06}
We find that, in contradistinction to CSG's Thomas-Fermi approach, partly filled LLs appear as a mixture of clusters of locally full and locally empty LLs. Stripes of nearly constant half-odd filling emerge at the boundaries of these clusters at higher LLs. We identify this behavior as a consequence of a $\nu$-dependent exchange-enhanced $g$-factor. The existence of an exchange-enhanced $g$-factor seems incompatible with a lateral \emph{smoothly} varying carrier density across CSs of a width that is clearly larger than the width of the incompressible stripes as obtained by models based on Thomas-Fermi screening alone.\cite{FogS95} 
These result demonstrate that the IQH regime is dominated by many-particle physics that seems to acts towards re-establishing the behavior expected for non-interacting single electrons --- as often assumed in early models of the IQH effect.  

We note that an extension of the CSG model for spin-split LLs was considered in Ref.\ \onlinecite{FogS95} in which also possible effects of the exchange interaction were discussed. A local filling factor dependence for the case of high-mobility heterostructures at moderate densities was assumed to lead to  "narrow strips of the compressible liquid, where $\nu(\mathbf{r})$ is half-integer, remain metallic".\cite{FogS95} This is very reminiscent of the half-odd integer features discussed, e.g., in section \ref{half-odd}.
However, the "0"-"2" to "0"-"1"-"2" transition described for the edge states in Ref.\ \onlinecite{FogS95} (cp.\ Figs.\ 4 and 5 of Ref.\ \onlinecite{FogS95}) appears at variation with the local Hund's rule behavior found in our work. Rather, the schematic picture advocated in Ref.\ \onlinecite{FogS95} is reminiscent of our results for the self-consistent Hartree calculations (cp.\ Fig.\ \ref{fig-00-HH-HF-up} (c) and Fig.\ \ref{fig-00-HH-HF-full} (d)).
Nevertheless, this does not imply a non-resolvable contradiction with the CSG model and its extension. Our results are valid in relatively small systems of about half a micron width. The CSG approach applies on larger length scales for stripes of hundreds of nanometer width and it appears likely that the mesoscopic regime we are investigating here is already beyond the validity of the CSG approach. Perhaps the varying cluster numbers and sizes within our model average to a quasi continuously varying carrier density for laterally much larger structures than the typical cluster size.
Taken literally, this would imply that the smooth CSs according to CSG, may have an internal cluster like structure, which divides the whole smoothly looking CS into a dense network of transmitting half-odd integer stripes, separating clusters of full and empty LLs. Our model could therefore be interpreted as the internal (fine-) structure of the almost macroscopically wide CSs of CSG.

\section{\label{acknowledge}Acknowledgments}

We thank B.\ Shklovskii for discussions and for pointing out Ref.\ \onlinecite{FogS95} to us. We gratefully acknowledge funding via the Austrian Science Foundation FWF Project P19353-N16 and provision of computing resources through the MidPlus Regional HPC Centre (EP/K000128/1) as well as the University of Warwick's Centre for Scientific Computing. UK research data statement: Data accompanying this publication is available at Ref.\ \onlinecite{QHsupplement2}.


\bibliographystyle{prsty}\bibliography{Mendeley_SGM-QH}

\ifNOSUP\end{document}\else%

\clearpage\newpage
\setcounter{figure}{0}
\setcounter{table}{0}
\def\thefigure{S\arabic{figure}}
\def\thetable{S\arabic{table}}
\setcounter{page}{1}
\pagestyle{plain}

\section*{Supporting Information}

{\center
\textbf{Manifestation of many-body interactions in the integer quantum Hall effect regime}\\[2ex]

\noindent%
Josef Oswald$^{1}$, Rudolf A R\"{o}mer$^{2,3}$\\[2ex]

$^1${Physics Institute, University of Leoben, Austria, Josef.Oswald@unileoben.ac.at}\\
$^2${Department of Physics and Centre for Scientific Computing, University of Warwick, Coventry, CV4 7AL, UK, R.Roemer@warwick.ac.uk}\\
$^3${Department of Physics and Optoelectronics, Xiangtan University, Xiangtan 411105, Hunan, China}\\
}
\hspace*{2ex}

\begin{figure}[b]
\includegraphics[width=0.44\textwidth]{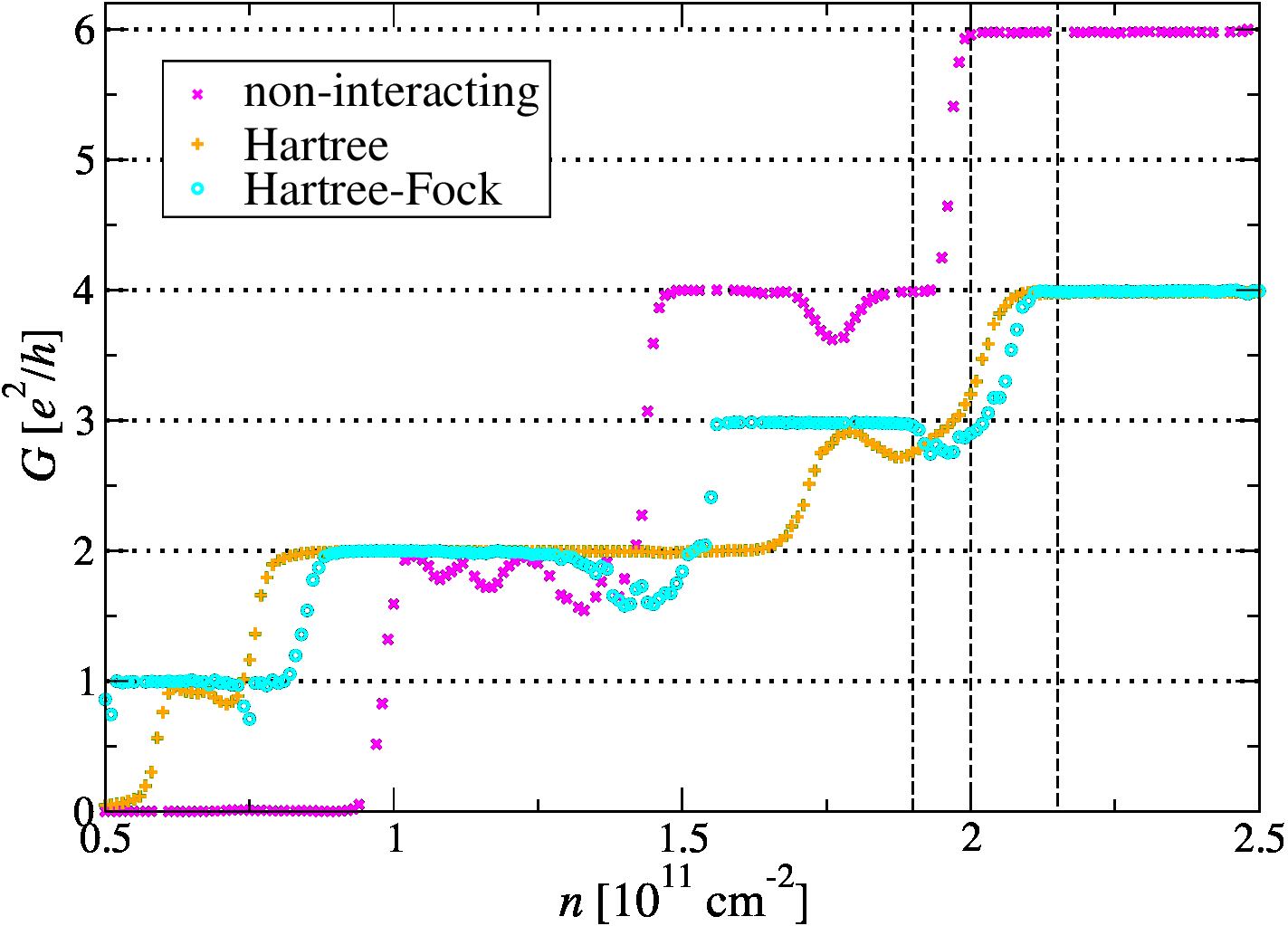}
\caption{\label{fig-Gxx_ns_00-HH-HF}\label{fig-conductance-density-00-HH-HF}
Two-point conductance $G$ versus carrier density $n$ at fixed magnetic field of $B=3$ T for the non-interacting ($\times$), the Hartree-interacting ($+$) and the Hartree-Fock-interacting model ($\circ$). The horizontal dotted lines indicate integer multiples of $e^2/h$ while the vertical dashed lines indidcate the three density values of $n= 1.9, 2.0, 2.15$ ($\times 10^{11}$ cm$^{-2}$).}
\end{figure}

\begin{figure*}[b]
(a)\includegraphics[width=0.6\textwidth]{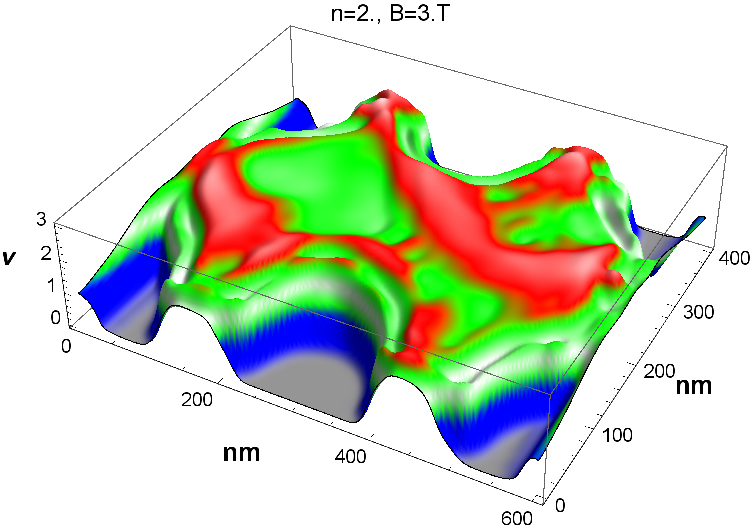} 
\begin{minipage}{0.32\textwidth}\vspace*{-50ex}
(b)\includegraphics[width=\textwidth]{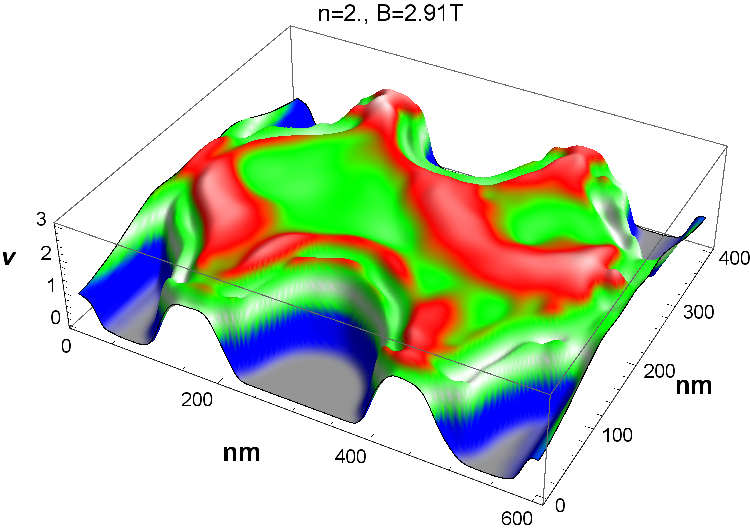}\\ 
(c)\includegraphics[width=\textwidth]{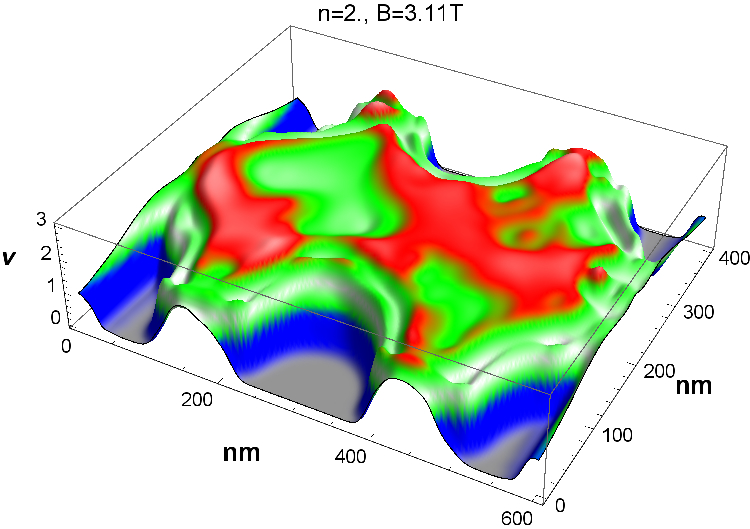}
\end{minipage}
(d)\includegraphics[width=0.3\textwidth]{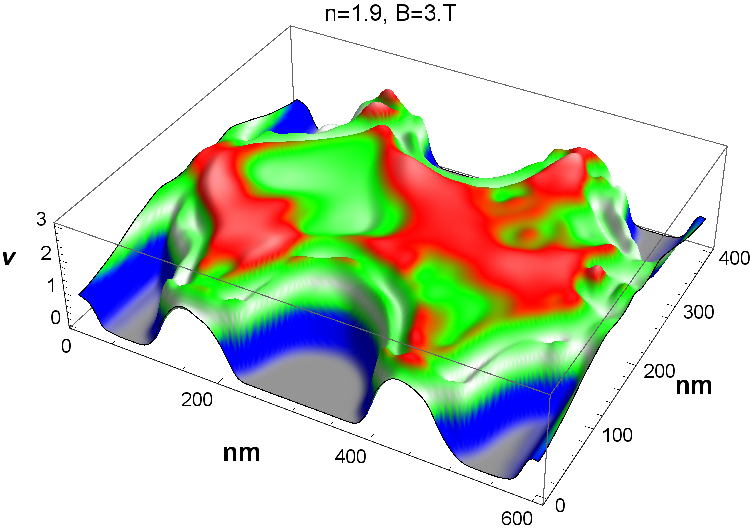}
(e)\includegraphics[width=0.3\textwidth]{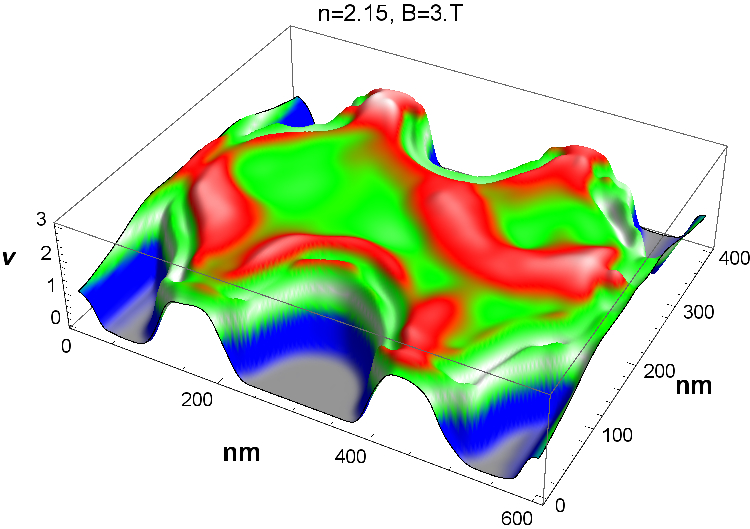}
(f)\includegraphics[width=0.3\textwidth]{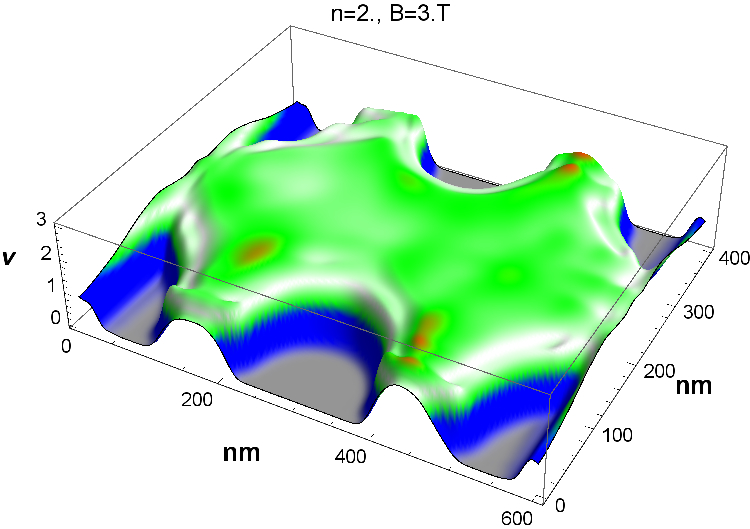}
\caption{\label{CDS1x3_sp2}\label{fig-fillingfactor-down} Lateral distribution of filling factor  $\nu_\downarrow$ corresponding to the same situation and parameters as in Fig.\ \ref{fig-fillingfactor-up}.
Colors are as in Fig.\ \ref{fig-fillingfactor-up}.
The filling factor range close to $\nu_\downarrow = 1.5$  is highlighted in light gray in order to identify possible stripes appearing close to the half filled top LL (there are none).}
\end{figure*}

\begin{figure*}[tb]
(a)\includegraphics[width=0.6\textwidth]{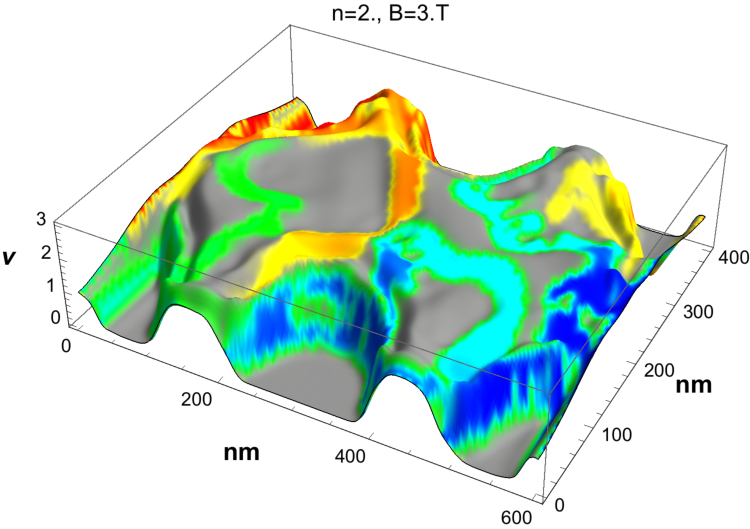}
\begin{minipage}{0.32\textwidth}\vspace*{-50ex}
(b)\includegraphics[width=\textwidth]{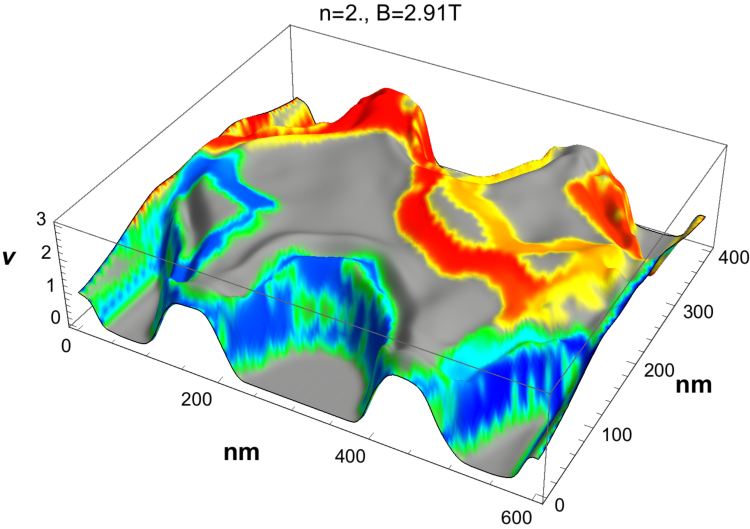}\\
(c)\includegraphics[width=\textwidth]{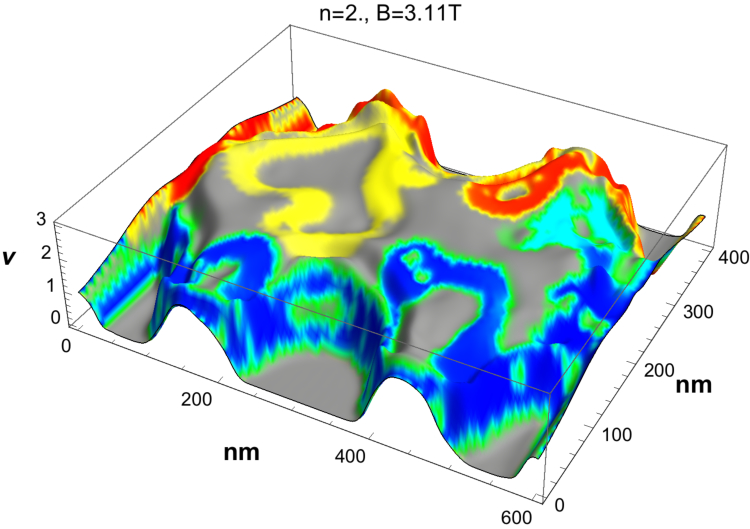}
\end{minipage}
(d)\includegraphics[width=0.3\textwidth]{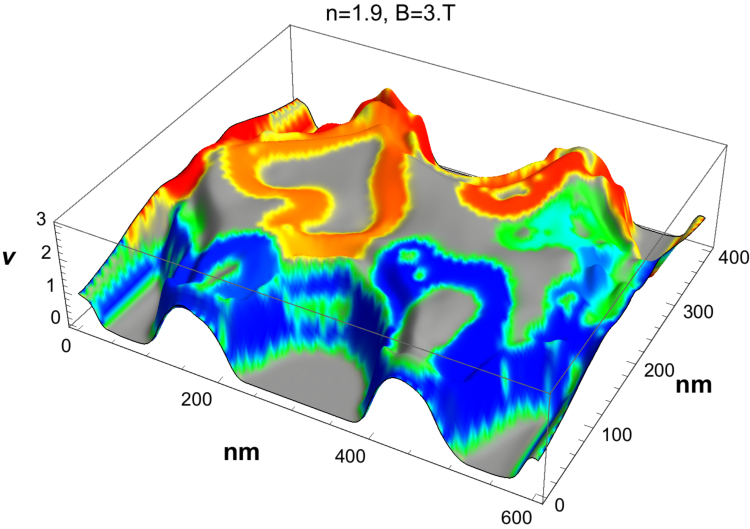}
(e)\includegraphics[width=0.3\textwidth]{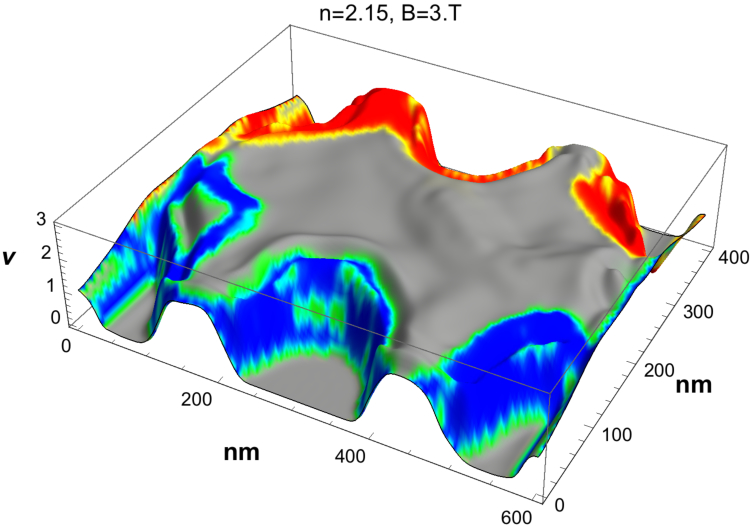}
(f)\includegraphics[width=0.3\textwidth]{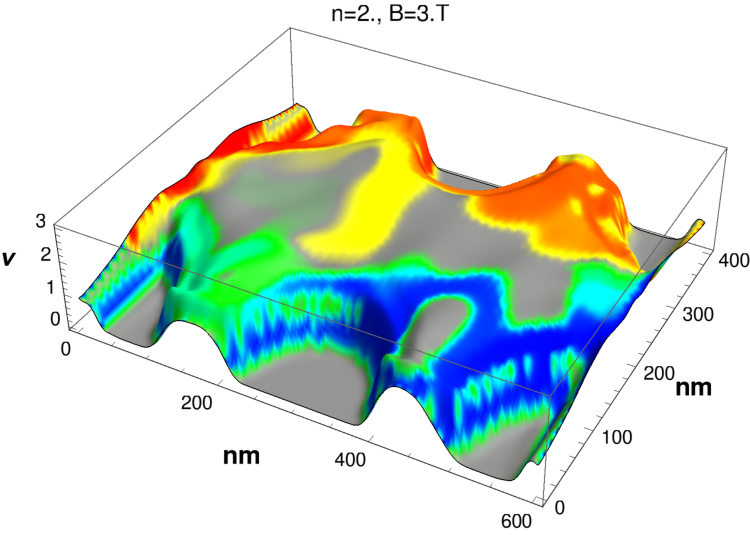}
\caption{\label{ECS2x3_sp2}\label{fig-chemicalpotential-down} Lateral non-equilibrium distribution of $\mu$-on-top-of-$\nu_\downarrow$ distribution of Fig.\ \ref{fig-fillingfactor-down} corresponding to the $\mu$-on-top-of-$\nu_\uparrow$ cases as in Fig.\ \ref{fig-chemicalpotential-up}. Colors are as in Fig.\ \ref{fig-chemicalpotential-up}.}
\end{figure*}

\begin{figure}[b]
\includegraphics[width=0.44\textwidth]{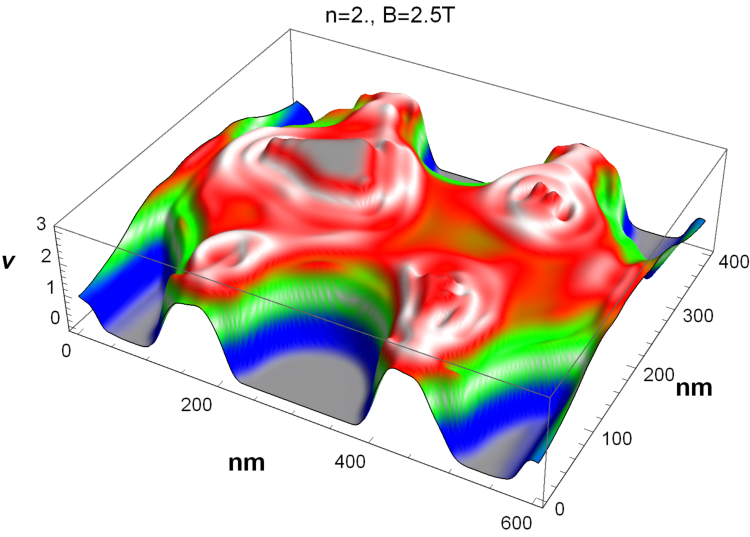}
\caption{\label{fig-fillingfactor-down-highLL}\label{HF_CD_B250_n200_S1030_sp2_select.jpg}
Spatial distribution of $\nu_\downarrow$ for the highest partly filled LL close to the $\nu = 4 \rightarrow 5$ plateau transition at $T=0$ 
with $B=2.5$ T and $n=2\times 10^{11}$ cm$^{-2}$. 
Colors are as in Fig.\ \ref{fig-fillingfactor-up} with dark gray denoting $\nu_\downarrow\in [4,5]$ and the light gray corresponding to $\nu_\downarrow= 1.5$ and $2.5$.
}
\end{figure}

\begin{figure*}[tb]
(a)\includegraphics[width=0.3\textwidth]{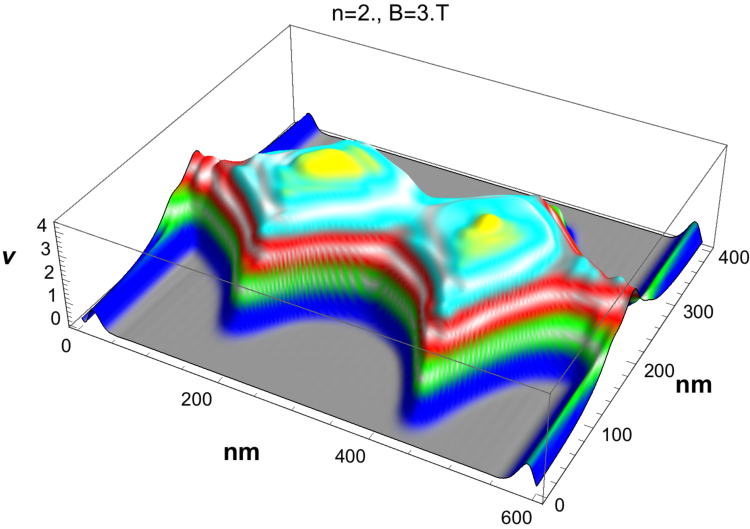}
(d)\includegraphics[width=0.3\textwidth]{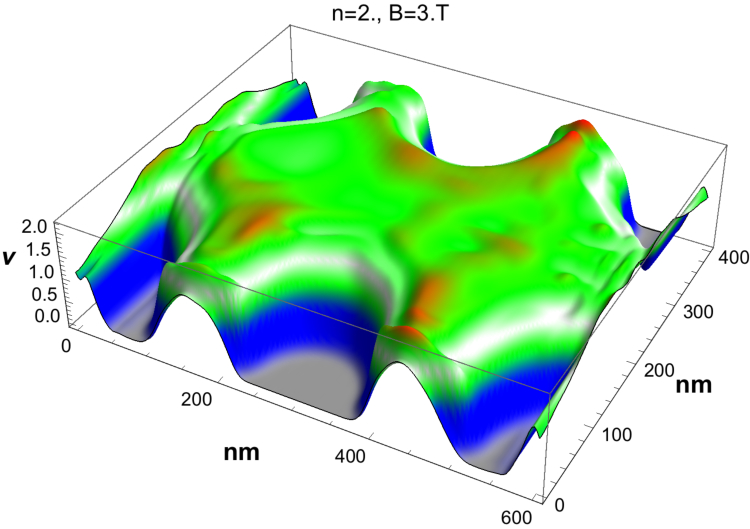} 
(g)\includegraphics[width=0.3\textwidth]{HF-CD_B300_n200_S1030_sp2.jpg}\\
(b)\includegraphics[width=0.3\textwidth]{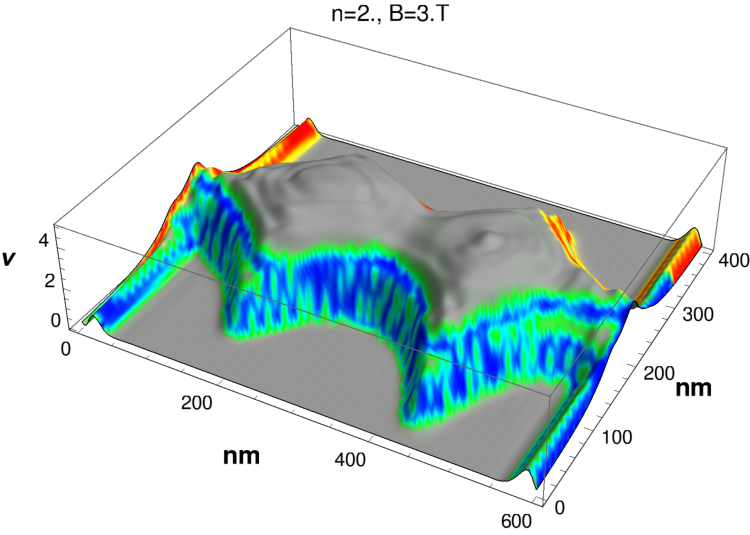}
(e)\includegraphics[width=0.3\textwidth]{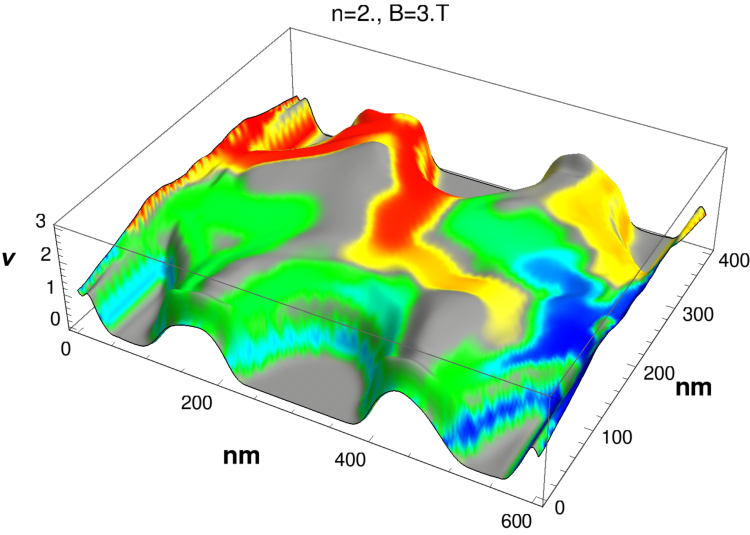} 
(h)\includegraphics[width=0.3\textwidth]{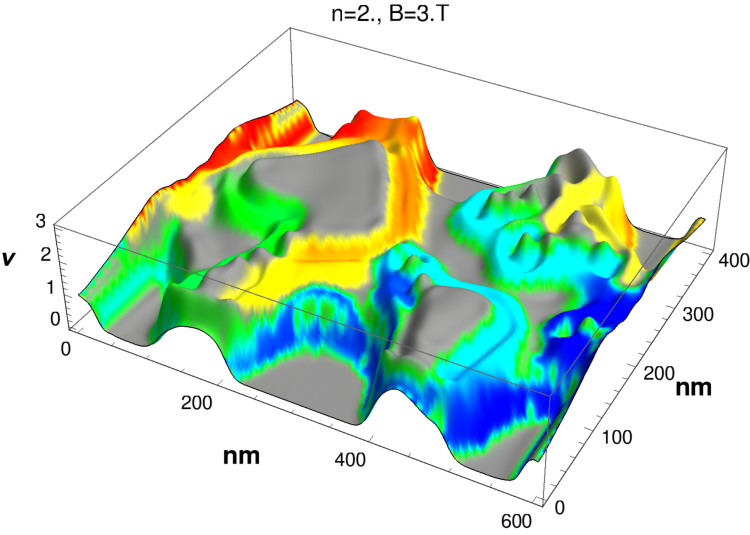}\\
(c)\includegraphics[width=0.3\textwidth]{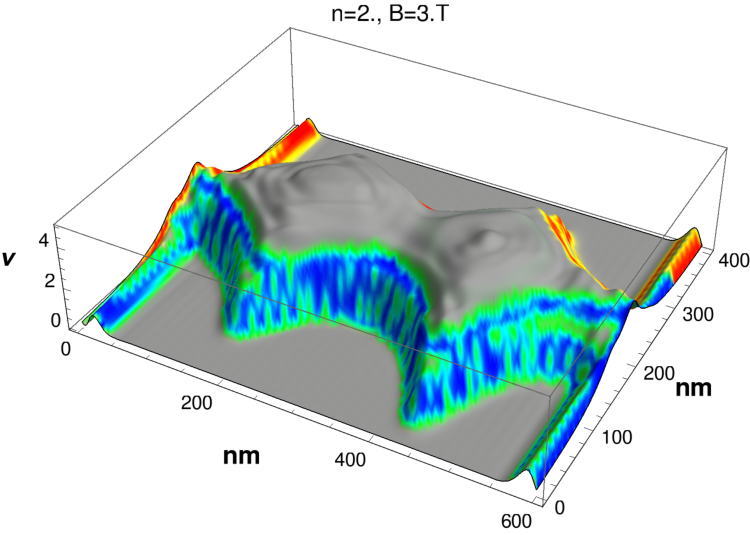}
(f)\includegraphics[width=0.3\textwidth]{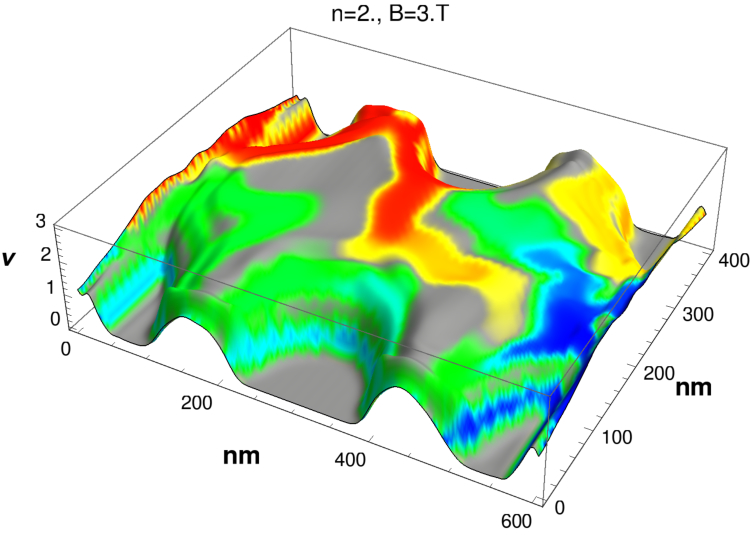}
(i)\includegraphics[width=0.3\textwidth]{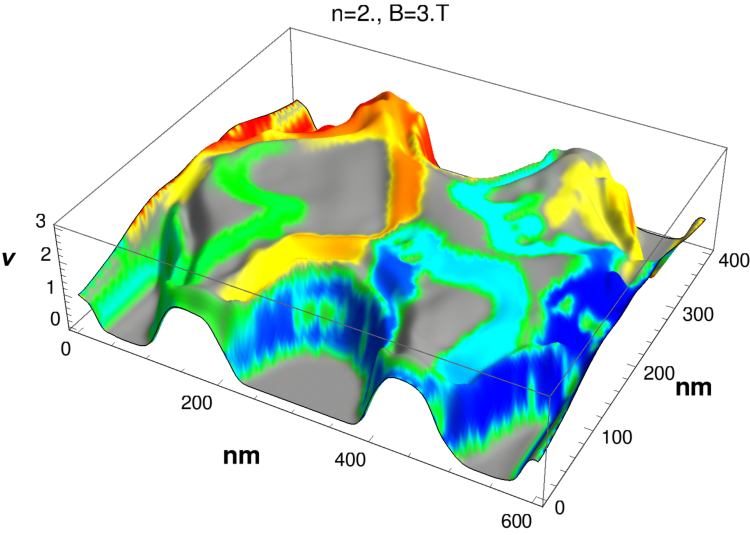}
\caption{\label{ECCD-n200-00-HH-HF_spX}\label{fig-00-HH-HF-full} Distribution of $\mu$ and local $\nu$ based on (a+b+c) the interaction-free single particle approximation, (d+e+f) the Hartree approximation  and (g+h+i) the Hartree-Fock approximation at $B=3$ T and $n=2\times 10^{11}$ cm$^{-2}$. The top row corresponds to the local distribution of $\nu_\downarrow$ (cp.\ Fig.\ \ref{fig-00-HH-HF-up} for $\mu$ on $\nu_\uparrow$), while the second and third rows denote $\mu$ on $\nu_\uparrow$ and $\nu_\downarrow$, respectively.
The colors are as in Figs.\ \ref{fig-fillingfactor-up} and \ref{fig-chemicalpotential-up}, as well as Fig.\ \ref{fig-00-HH-HF-up}.
}
\end{figure*}

Fig.\ \ref{fig-conductance-density-00-HH-HF} shows the density dependence of the conductance similar to Fig.\ \ref{fig-conductance-resistance-density}, but with results for the non-interacting and the solely Hartree-interacting cases included as well for comparison.
The other supplemental information provides the figures for the spin-down case corresponding to Figs.\ \ref{fig-fillingfactor-up} and \ref{fig-chemicalpotential-up}, that is, the spatial distribution of the filling factor $\nu_\downarrow$ in Fig.\ \ref{fig-fillingfactor-down} and the non-equilibrium chemical potential $\mu$ plotted on top of $\nu_\downarrow$ in Fig.\ \ref{fig-chemicalpotential-down}. 
Fig.\ \ref{fig-fillingfactor-down-highLL} shows a situation analogous to Fig.\ \ref{fig-fillingfactor-up}, but now for $\nu_\downarrow$ in a higher LL.
In addition, we supplement Fig.\ \ref{fig-00-HH-HF-up} for $\nu_\uparrow$ with its $\nu_\downarrow$ version as well as the corresponding $\mu$ distributions in Fig.\ \ref{fig-00-HH-HF-full} for the non-interacting, the Hartree-interacting and the full Hartree-Fock-interacting situation. 

\fi\end{document}